\documentclass[a4paper,10pt]{article}
\usepackage[position=t,singlelinecheck=off]{subfig}
\usepackage{amssymb,amsmath,graphics,graphicx,float,braket}
\usepackage{array,multirow,setspace}
\def\beq{\begin{equation}}
\def\eeq{\end{equation}}
\def\bea{\begin{eqnarray}}
\def\eea{\end{eqnarray}}
\def\nn{\nonumber}

\def\dm{\rho^{(\kappa)}_{\mbox{\fontshape{it}\scriptsize{n,m}}}}
\def\numdm{\rho^{(\kappa)}_{\mbox{\fontshape{it}\scriptsize{n$+$$\ell$,m$+$$\ell$}}}}

\makeatletter
\newcommand{\lvast}{\bBigg@{3}}
\newcommand{\svast}{\bBigg@{2}}
\newcommand{\vast}{\bBigg@{4}}
\newcommand{\Vast}{\bBigg@{5}}
\makeatother
\makeatletter
\def\@cite#1#2{${\mbox{#1\if@tempswa , #2\fi}}$}
\makeatother

\begin{document}
\thispagestyle{empty}
\begin{center}
\begin{LARGE}
\textsf{Nonclassicality and decoherence of 
	photon-added  squeezed coherent Schr\"{o}dinger kitten states in a Kerr medium}
\end{LARGE} \\

\bigskip\bigskip
R. Chakrabarti$^{\dagger}$ and V. Yogesh$^{\ddagger}$
\\
\begin{small}
\bigskip
\textit{
 $^{\dagger}$Chennai Mathematical Institute, H1 SIPCOT IT Park, \\Siruseri, Kelambakkam 603 103, India\\
$^{\ddagger}$ Department of Theoretical Physics, 
University of Madras, \\
Maraimalai Campus, Guindy, 
Chennai 600 025, India \\}
\end{small}
\end{center}

\vfill
\begin{abstract}
We study the nonclassicality of the evolution of a superposition of an arbitrary number of photon-added squeezed coherent Schr\"{o}dinger
 cat states in a nonlinear Kerr medium. The nonlinearity of the medium gives rise to the periodicities of  the  quantities such as the Wehrl entropy $S_{Q}$ and the negativity $\delta_{W}$ of the $W$-distribution, and a series of local minima of these quantities arise at the rational submultiples of the said period. At these local minima the evolving state coincides with the transient Yurke-Stoler type of photon-added squeezed kitten states, which, for the choice of the phase space variables reflecting  their macroscopic nature, show extremely short-lived behavior. Proceeding further we provide the closed form  tomograms, which furnish the alternate description of these short-lived states. The increasing complexity in the kitten formations induces  more number of interference terms that trigger more quantumness of the          
corresponding states. The nonclassical depth of the photon-added squeezed kitten states are observed to be of maximum possible value. 
Employing the Lindblad master equation approach we study the amplitude and the phase damping models for the initial state considered here. In the phase damping model the nonclassicality is not completely erased even in the long time limit when the dynamical quantities, such 
as the negativity $\delta_{W}$ and the tomogram, assume nontrivial asymptotic values.   
\end{abstract} 
 
\newpage
\setcounter{page}{1}
\section{Introduction}

The nonclassical states play an important role in elucidating the fundamental principles of quantum mechanics and are expected to 
provide advantages in various quantum information protocols. These states are  characterized by non-Gaussian features in their 
phase space quasiprobability distributions which usually are realized due to Kerr-type nonlinear dispersive  medium exhibiting 
the third order light-matter interaction [\cite{S1984}]. It has been recently noticed [\cite{MT1994}] that  a quantum mechanical 
resonator, whose tiny vibrations are controlled by the radiation pressure of the cavity field,  mimics a Kerr oscillator when the cavity field is driven by a coherent light.  Employing the Born-Oppenheimer approximation for the optomechanical system  an effective 
Hamiltonian that leads to a nonlinear Kerr effect has been constructed [\cite{GILSN2009}]. Moreover, the optomechanical system has been found [\cite{GILSN2009}] to induce a squeezing effect in the intensity spectrum of the cavity field. The optical Kerr effect has been 
employed [\cite{IHY1985}] towards a quantum nondemolition measurement of the photon number, and its accuracy, which is linked with 
the imposed phase noise via the uncertainty principle, has been studied. It has also been observed [\cite{DA2007}] that under the 
condition of electromagnetically induced transparency the Kerr effect can make a very significant contribution to
the group velocity of the slow light propagation through a four-level $N$-type atomic system. 

\par

The propagation of quantum states of light in a Kerr medium has been extensively studied. In particular, it has been observed 
[\cite{MTK1990}-\cite{MBWI2001}] that an initial coherent state therein evolves, at certain specified times, to  superposition of 
coherent states popularly known as Schr\"{o}dinger kitten states that display macroscopic characteristics for a large value of the 
relevant complex amplitude in the phase space. Assuming a squeezed coherent state as input for a nonlinear Kerr medium it has 
been found [\cite{TAC1993}]
that its evolution provides a finite superposition of squeezed kitten states at certain instants. The squeezed states of the harmonic oscillator are, however, of much experimental focus as they involve the reduction of the fluctuations in one quadrature variable below the ground state uncertainty. These states play a key role in quantum metrology towards improving the sensitivity of the interferometers [\cite{Goda2008}], and are of crucial importance in the recent gravitational wave detection via high-power laser interferometers [{\cite{AAAA2013}]. The squeezed states are also important in the continuous variable quantum key
distribution protocols [\cite{LSKWL2001}], and they impart [\cite{UF2011}] an enhancement in comparison to their coherent analogs. Moreover, they have recently been used as sensitive detectors for photon scattering recoil events at the single
photon level [\cite{HLJGBR2013}].

\par

One way of enhancing the nonclassicality of a quantum light state is to add (subtract) a finite number of photons to (from) 
another suitable light state. The sub-Poissonian character of the fields in states produced by the addition of photons to a coherent state has been studied in [\cite{AT1991}]. The nonclassical properties of the single photon-subtracted squeezed vacuum as well as the coherent states have been observed [{\cite{BA2007}, \cite{WMF2012}] specifically in terms of the sub-Poissonian statistics and the corresponding phase space quasiprobability distributions. Experimental generation of a single photon-added thermal state has been achieved [\cite{ZPB2007}] and its quantumness has 
been studied by constructing the negativity of the corresponding Wigner distribution. Using a single photon interferometer towards 
realizing the coherent superpositions of two alternate sequences of photon addition and subtraction, an experimental test of the bosonic commutation relation between the annihilation and creation operators  has been implemented  [\cite{KJZPB2008}, \cite{PZKB2007}]. The possibility of arbitrarily adding and subtracting photons to and from a light field may allow us to engineer  customized fruitful
quantum states. Using a conditional addition or subtraction of photons on the Gaussian entangled beams an enhancement of the entanglement has been observed [\cite{ODTG2007}, \cite{NGSC2012}]. 

\par

In the present work we investigate the evolution of an arbitrary number of photon-added squeezed coherent Schr\"{o}dinger cat state in a nonlinear medium. 
Our preferred tools in this regard are the quasiprobability distributions in the phase space. The diagonal $P$-representation 
[\cite{S1963}, \cite{G1963}] in a coherent state basis is highly singular in nature pointing towards the underlying nonclassicality 
of the evolving state. The Wigner $W$-distribution [\cite{W1932}] offers a link between the classical and quantum dynamics, as unlike a true classical probability distribution it assumes negative values providing a quantitative  indicator of the extent of nonclassicality. For this purpose the negative volume in the phase space of the $W$-distribution $\delta_{W}$ [\cite{KZ2004}] has been employed.  
The  suitably smoothed nonnegative Husimi Q-function [\cite{H1940}] in the phase space facilitates the construction of the semiclassical Wehrl entropy [\cite{W1978}]. For the nonlinear Kerr medium the dynamical quantities such as the Wehrl entropy and negativity $\delta_{W}$ 
follow a periodic structure, and, in particular, exhibit a series of local minima at the rational submultiples of the period. Transient
photon-added squeezed Schr\"{o}dinger kitten states arise precisely at these minima. The optical tomogram [\cite{VR1989}-\cite{LR2009}] describes the probability distribution of the rotated quadrature phase, and presents  an equivalent description to the one offered by the quasiprobability functions. We study the optical tomogram of the time evolved state, and obtain closed form expressions  for the 
tomograms  corresponding  to the transient photon-added  squeezed kitten formations. Another way of studying the nonclassicality of the state is advanced by the nonclassical depth [{\cite{L1991}, \cite{L1992}] that smears a highly singular $P$-representation with 
a Gaussian function possessing a variable dispersion so that the smoothed phase space distribution $R(t)$, for a restricted choice of the dispersion parameter, assumes nonnegative values. The nonclassical depth has recently been used [{\cite{MKV2016}] to assess the nonclassicality of superposition of quantum states prepared by permutations of nondemolition quadrature experiments and single quanta addition or subtraction. It has also been applied [\cite{HWLZ2016}] in  quantitatively determining the nonclassicality of the non-Gaussian states prepared by utilizing the multiphoton catalysis with coherent state input.
We obtain  analytical expressions of the $R(t)$-distribution for the short-lived kitten-like structures, and numerically find that these states are associated with the maximum possible nonclassical depth. Incorporating the decoherence of the initial state via the Lindblad master equation [\cite{L1976}, \cite{GFVKS1978}], we study both the amplitude damping and the phase damping processes. The analytical solution of the master equation for the example of amplitude damping implies that the system asymptotically decays  to its ground state. In contrast, the mechanism of the phase dissipation process erases the off-diagonal terms of the density matrix while leaving the diagonal entries invariant. The phase damping model, therefore, partially retains nonclassicality even in the long time limit. Employing the evolution equation we analytically obtain the asymptotic steady state values of the dynamical quantities such as the quasiprobability distributions and the tomogram corresponding to the phase damping model. 
 
\section{Photon-added squeezed coherent cat state in a Kerr medium and phase space distributions}
\label{PASC}
 The Hamiltonian of a  single field mode in a nonlinear Kerr medium [\cite{TMK1991}] reads in natural units ($\hbar=1$) as follows:   
\beq
H = \omega \, a^{\dagger} a + \lambda \, a^{\dagger\,2}a^{2},
\label{KerrH}
\eeq
where the oscillator of frequency $\omega$ is described by the annihilation and creation operators 
($a, a^{\dagger}| \hat{n} \equiv a^{\dagger} a$), and the coupling constant $\lambda$ corresponds 
to the third-order susceptibility of the Kerr medium. 
We study the evolution of the initial ($t=0$) multiple ($\kappa$) photon-added squeezed coherent Schr\"{o}dinger cat state given by 
\beq
\ket{\psi^{(\kappa)}_{}(0)} = \mathcal{N}^{(\kappa)} 
\, a^{\dagger \kappa} \, \left( \ket{\xi, \alpha} + \mathrm{c} \, 
\ket{ \xi,-\alpha} \right),  \quad |\alpha\rangle = \mathrm{D}(\alpha)|0 \rangle,\,  
\ket{ \xi, \alpha} =  \mathrm{S}(\xi) \mathrm{D}(\alpha) \ket{0}, \, \mathrm{c} \in \mathbb{C}.
\label{initial-state-mpa}
\eeq
The displacement and the squeezing operators given in (\ref{initial-state-mpa}) read, respectively, as
$\mathrm{D}(\alpha)= \exp(\alpha a^{\dagger}-\alpha^*a), \, \mathrm{S}(\xi)= \exp((\xi a^{\dagger}{^2}-\xi^{*} a^2 )/2),\, 
\alpha = |\alpha| \exp(i \,\theta),\, \xi = r \exp(i\,\vartheta),\; \alpha, \xi \in \mathbb{C}$.
The squeezing operator maintains the following unitary transformations:
\beq
\mathrm{S}^{\dagger}(\xi)\, a \,\mathrm{S}(\xi)= \mu\, a + \nu\, a^{\dagger}, \;\; \mathrm{S}^{\dagger}(\xi)\, a^{\dagger} \,
\mathrm{S}(\xi)=  \mu\, a^{\dagger} +  \nu^{*}\, a,\quad \mu=\cosh r,\;\;\nu= \exp(i \vartheta)\,\sinh r.
\label{S-unitary}
\eeq
The squeezed coherent state in the number state basis may be written [\cite{SZ2001}] as
\beq
\ket{\xi,\alpha}=\exp \Big(-\dfrac{|\alpha|^{2}}{2}-\dfrac{\alpha^{2}\nu^{*}}{2\mu} \Big)
\sum_{n=0}^{\infty} \dfrac{i^{n}}{\sqrt{n!\mu}}
\left( \dfrac{\nu}{2\mu}\right)^{\tfrac{n}{2}} \mathrm{H}_{n}\left(-i\, \dfrac{\alpha}{\sqrt{2 \mu \nu}}\right) \ket{n},
\label{squeezed-coherent-state}
\eeq
where the Hermite polynomials are described by the generating function: 
$\exp (2\,\mathsf{x} \mathsf{t} - \mathsf{t}^{2}) = \sum_{n=0}^{\infty} \tfrac{\mathrm{H}_{n}(\mathsf{x})\, \mathsf{t}^{n}}{n!}$.
Similarly, the multiple photon-added squeezed coherent state in the number state basis reads  $a^{\dagger \kappa} \ket{\xi,\alpha}
 \equiv \displaystyle{\sum_{n=\kappa}^{\infty}} \mathcal{A}_{n,\kappa}( \xi,\alpha) |n \rangle$,
where the coefficients are given by
\beq
\mathcal{A}_{n,\kappa}( \xi,\alpha) = 
\exp \Big(-\dfrac{|\alpha|^{2}}{2}-\dfrac{\alpha^{2}\nu^{*}}{2\mu} \Big)
\dfrac{i^{n-\kappa}\,\sqrt{n!}}{(n-\kappa)!\,\sqrt{\mu}} \left( \dfrac{\nu}{2\mu} \right)^{\tfrac{n-\kappa}{2}} 
\mathrm{H}_{n-\kappa}\left(-i\, \dfrac{\alpha}{\sqrt{2 \mu \nu}}\right).
\label{A_nk}
\eeq
The parity property of the Hermite polynomials $\mathrm{H}_{n}(-\mathsf{x}) = (-1)^{n}\,\mathrm{H}_{n}(\mathsf{x})$ relates the above coefficients as follows:
$\mathcal{A}_{n,\kappa}( \xi,- \alpha) = (-1)^{n- \kappa}\,\mathcal{A}_{n,\kappa}( \xi,\alpha)$. The normalization constant for 
the initial state (\ref{initial-state-mpa}) reads
\beq
\mathcal{N}^{(\kappa)} \!  = \left\lgroup\sum_{n=\kappa}^{\infty} \left\lgroup 1 + |\mathrm{c}|^{2} + 2 \, (-1)^{n- \kappa}\, 
\mathrm{Re} (\mathrm{c})\right\rgroup \,|\mathcal{A}_{n,\kappa}( \xi,\alpha)|^{2}\right\rgroup^{-\tfrac{1}{2}},
\label{norm-mpa}
\eeq
where the infinite sums relevant for the above construction may be explicitly evaluated: 
\bea
\sum_{n=\kappa}^{\infty}
\label{sum_1_mpa}
|\mathcal{A}_{n,\kappa}( \xi,\alpha) |^{2} &=& \kappa! \sum_{p=0}^{\kappa}  \sum_{\ell=0}^{p} (-1)^{p} \binom{\kappa}{p} 
\dfrac{|\nu|^{2 p - \ell}}{2^{p}\,\ell!\,(p-\ell)!}\times \nn \\
&& \times\;
\mathrm{H}_{\ell}\left( \dfrac{i\alpha}{\sqrt{2\mu \nu}}\right) 
\mathrm{H}_{2p-\ell} \left( i\alpha \sqrt{\dfrac{\nu^{*}}{2\mu}} +i \alpha^{*} \sqrt{\dfrac{\mu}{2\nu^{*}}} \right),\\
\! \! \! \! \sum_{n=\kappa}^{\infty}  (-1)^{n -\kappa}
|\mathcal{A}_{n,\kappa}( \xi,\alpha)|^{2} 
&=&  \exp(-2|\alpha|^{2}) \; \kappa! \sum_{p=0}^{\kappa} \sum_{\ell=0}^{p} (-1)^{p} \binom{\kappa}{p} 
\dfrac{|\nu|^{2 p - \ell}}{2^{p}\,\ell!\,(p-\ell)!}\times \nn \\ 
&&  \times \; 
\mathrm{H}_{\ell}\left( \dfrac{i\alpha}{\sqrt{2\mu \nu}}\right) 
\mathrm{H}_{2p- \ell} \left( i\alpha \sqrt{\dfrac{\nu^{*}}{2\mu}} - i \alpha^{*} \sqrt{\dfrac{\mu}{2\nu^{*}}} \right).
\label{sum_2_mpa}
\eea
To compute the  lhs of (\ref{sum_1_mpa}, \ref{sum_2_mpa}) we have employed the following identity of the 
Hermite polynomials:
\bea
\sum_{n=0}^{\infty}\dfrac{(n+\kappa)!\, \mathsf{t}^{n}}{2^{n}(n!)^{2}} \, \mathrm{H}_{n}(\mathsf{x}) \,\mathrm{H}_{n}(\mathsf{y}) &=& 
\dfrac{\kappa!}{\sqrt{1-\mathsf{t}^{2}}} \;
\mathcal{E}_{\mathsf{t}}(\mathsf{x},\mathsf{y}) 
\sum_{p=0}^{\kappa} \sum_{\ell=0}^{p} (-1)^{p} \binom{\kappa}{p} \times \nn\\ 
&&\times\; \dfrac{1}{2^{p}\, \ell!(p-\ell)!}\,\left(\dfrac{\mathsf{t}}{\sqrt{1-\mathsf{t}^{2}}}\right)^{2p -\ell}\, 
\mathrm{H}_{\ell}(\mathsf{x}) \, \mathrm{H}_{2p-\ell}\left(\mathsf{\dfrac{\mathsf{t x}-\mathsf{y}}{\sqrt{1-\mathsf{t}^{2}}}}\right),
\label{Hermite_identity_mpa}
\eea
where the exponential factor reads: $\mathcal{E}_{\mathsf{t}}(\mathsf{x},\mathsf{y})=\exp\left(-\dfrac{(\mathsf{tx})^{2}-2\mathsf{txy}+
(\mathsf{ty})^{2}}{ 1-\mathsf{t}^{2}} \right)$. 
The time evolution of the initial state is given by,
\beq
\ket{\psi^{(\kappa)}_{}(t)}=\mathcal{N}^{(\kappa)}\sum_{n=\kappa}^{\infty}
\left(1+(-1)^{n-\kappa} \,\mathrm{c}\right) \mathsf{A}_{n,\kappa}(t)\, |n\rangle, \;\; 
\mathsf{A}_{n,\kappa}(t) = \mathcal{A}_{n,\kappa} (\xi,\alpha)\, \exp(-i ((\omega-\lambda)n+ \lambda n^{2})t)
\label{state-t}
\eeq
and the corresponding density matrix assumes the form 
\beq
\rho^{(\kappa)}(t)\equiv \ket{\psi^{(\kappa)}_{}(t)}\bra{\psi^{(\kappa)}_{}(t)}
=   \sum_{n,m=\kappa}^{\infty}
\dm(t) \ket{n}\bra{m},
\label{density_matrix_mpa}
\eeq
where the  matrix elements read
\beq
\dm(t)\equiv \braket{n|\rho^{(\kappa)}(t)|m}= \left(\mathcal{N}^{(\kappa)}\right)^{2}
\left(1+(-1)^{n-\kappa} \mathrm{c}\right)\, \left(1+(-1)^{m-\kappa} 
\mathrm{c}^{*}\right)\, \mathsf{A}_{n,\kappa}(t) \,\mathsf{A}_{m,\kappa}^{*}(t).
\label{den_nm}
\eeq
Owing to the relation (\ref{norm-mpa}) the normalization condition $ \mathrm{Tr} \rho^{(\kappa)}(t)=1$ is maintained.
\subsection{The diagonal Sudarshan-Glauber $P$-representation}
\label{Sudarshan}
It is well-known [\cite{S1963}, \cite{G1963}] that the overcompleteness of the coherent states admits the 
 construction of a radiation field density matrix employing the diagonal  projectors: 
\beq
\rho = \int P(\beta,\beta^{*}) \ket{\beta}  \bra{\beta}  \mathrm{d}^{2}\beta, \qquad
\int P(\beta, \beta^{*}) \mathrm{d}^{2}\beta = 1,
\label{P_def}
\eeq
where the normalized phase space distribution $P(\beta,\beta^{*})$ may be extracted  [\cite{M2009}] by inverting the defining property (\ref{P_def})
as follows: 
\beq
P(\beta,\beta^{*})= \dfrac{\exp(|\beta|^{2})}{\pi ^{2}} 
\int \bra{-\gamma}\rho \ket{\gamma} \,\exp(|\gamma |^{2})\, 
\exp(\beta  \gamma^{*} - \beta^{*} \gamma)\, \mathrm{d}^{2}\gamma.
\label{P_evaluation}
\eeq
The  density matrix (\ref{density_matrix_mpa}) now yields  the diagonal representation $P(\beta,\beta^{*})$ as a series sum
\beq
P^{(\kappa)}(\beta,\beta^{*},t)= \exp\left(|\beta|^{2}\right)
\sum_{n,m=\kappa}^{\infty} \dfrac{(-1)^{n+m}}{\sqrt{n!m!}}\,  \dm(t)
\left(\frac{\partial}{\partial \beta}\right)^{n} \left(\frac{\partial}{\partial \beta^{*}}\right)^{m}\, \delta^{(2)}(\beta)
\label{P_equation_mpa}
\eeq
that includes all derivatives of the delta function. The highly singular nature of the distribution (\ref{P_equation_mpa}) provides 
a clear signature of the nonclassicality of the evolving state. 
 The $P$-representation (\ref{P_equation_mpa}) obeys the normalization condition. Other phase space distributions   
such as Wigner $W$-distribution and the Husimi $Q$-function 
may be procured from the singular $P$-representation via appropriate smoothing operations performed by the Gaussian kernels.

\subsection{The Wigner $W$-distribution}
\label{Wigner}
An arbitrary oscillator density matrix admits the normalized Wigner quasiprobability distribution [\cite{W1932}] 
 defined via the displacement operator as
\beq
W(\beta, \beta^{*}) = \dfrac{1}{\pi ^{2}} \int \mathrm{Tr}(\mathrm{D}(\gamma)\rho) \exp\Big( \beta \gamma^{*}
- \beta^{*} \gamma \Big) \mathrm{d}^{2}\gamma,\quad \int W(\beta,\beta^{*})\; \mathrm{d}^{2}\beta = 1.
\label{W_def}
\eeq
Its alternate series representation [\cite{MK1993}], however,  provides an easy computational procedure: 
\beq
W (\beta, \beta^{*}) = \dfrac{2}{\pi} \sum_{\ell=0}^{\infty} (-1)^{\ell} \bra{\beta,\ell} 
\rho \ket{\beta,\ell}, \qquad \ket{\beta, \ell} = \mathrm{D}(\beta) \ket{\ell}.
\label{W_series}
\eeq
The evolution of the Wigner distribution may now be computed by substituting 
the density matrix (\ref{density_matrix_mpa}) in the series expansion (\ref{W_series}):
\bea
W^{(\kappa)}(\beta,\beta^{*};t) = \!\!\!\! &  \dfrac{2}{\pi} & \!\!\!\! 
\, \exp(-2|\beta|^{2})
\sum_{n,m=\kappa}^{\infty} \! \! \!  \dfrac{(2\beta^{*})^{n} (2\beta)^{m}}{\sqrt{n!m!}} 
 {}_2F_0\Big( -n,-m;\phantom{}_{-} ; -\dfrac{1}{4|\beta|^{2}}\Big) \dm(t) .
\label{W_function_mpa}
\eea
In the construction (\ref{W_function_mpa})  the identity [\cite{M1939}] 
\bea
\sum_{\ell=0}^{\infty} \frac{ (-\mathsf{t})^{\ell}}{\ell!}\, {}_2F_0\Big( -n,-\ell;\phantom{}_{-} ; -\frac{1}{\mathsf{t}}\Big) 
\,{}_2F_0\Big( -\ell,-m;\phantom{}_{-} ; -\dfrac{1}{\mathsf{t}}\Big) = 
2^{n+m} \exp(-\mathsf{t}) \; {}_2F_0\Big( -n,-m;\phantom{}_{-} ; -\frac{1}{4\mathsf{t}}\Big)
\label{IdentityHyper}
\eea
resulting from the bilinear generating function of the Charlier polynomials has been used.  
The hypergeometric sum is defined as $ \displaystyle{ {}_2F_0(\mathsf{x},\mathsf{y};\phantom{}_{-}; \tau)
	= \sum_{\ell=0}^{\infty} (\mathsf {x})_{\ell} (\mathsf{y})_{\ell}\, \frac{\tau^{\ell}}{\ell!},\,
	(\mathsf {x})_{\ell} = \prod_{j = 0}^{\ell - 1}(\mathsf {x} + j) }$. 
The series sum  (\ref{W_function_mpa}) also follows from the integral structure [\cite{SZ2001}]
\beq
W(\beta,\beta^{*}) = \frac{2}{\pi}\int P(\gamma, \gamma^{*})\, \exp(-2|\beta-\gamma|^{2}) 
\;\mathrm{d}^{2}\gamma
\label{P_int_W}
\eeq 
that manifests  the convolution of a smoothing Gaussian
kernel of variance $1/2$ on the phase space with the diagonal $P$-representation. The distribution property 
\bea
\int \!\!\!\!\!\!\! && \!\!\!\!\!\!\!\exp \Big( -\dfrac{|\beta - \gamma|^{2}}{\sigma}+|\gamma |^{2}\Big) 
\Big( -\dfrac{\partial}{\partial \gamma^{*}} \Big)^{n} 
\Big( -\dfrac{\partial}{\partial \gamma} \Big)^{m} \delta^{(2)}(\gamma)\, \mathrm{d}^{2}\gamma\nn\\
&=& \sigma^{-(n+m)} \beta^{n} \beta^{*m}  \exp \left(-\dfrac{|\beta|^{2}}{\sigma} \right) \,{}_2F_0\Big( \! \! -n,-m;\phantom{}_{-};
-\frac{\sigma(1-\sigma)}{|\beta|^{2}}\Big)
\label{Int_hyper}
\eea
for the choice $\sigma=1/2$ facilitates the integration in (\ref{P_int_W}) allowing us to reproduce the quasiprobability density 
(\ref{W_function_mpa}). This acts as a consistency check for our derivation. 
The expression (\ref{W_function_mpa}) reveals that corresponding to the choices ($\mathrm{c}=\pm 1$)
 of the even and odd combinations of the initial state (\ref{initial-state-mpa}) the
$W$-distribution assumes a parity symmetric form:
\beq
W^{(\kappa)}(\beta,\beta^{*};t)= W^{(\kappa)}(-\beta,-\beta^{*};t),
\label{W_symmetry}
\eeq
This property is responsible for the generation of only even number of kitten states 
(Fig. \ref{fig_kittens_mpa} $\mathsf{c}_{1}, \mathsf{c}_{2}, \mathsf{c}_{3}$) described in Sec. \ref{PASC} \ref{kittenTom}. 
On the other hand, for the choice $\mathrm{c}=\pm i$ \textit {\`{a} la } the Yurke-Stoler type of states [\cite{YS1986}] the symmetry (\ref{W_symmetry}) is violated
causing the production of both the even and odd number of kitten states. 
\par
It is well-known that the negative values of the  
 Wigner quasiprobability $W$-distribution act as an indicator of the nonclassicality of the state. The negative volume of the 
 $W$-distribution in the phase space 
 has been proposed [\cite{KZ2004}] as a quantitative measure of the property:
\beq 
\delta_{W} = \int |W(\beta, \beta^{*})| \mathrm{d^2}\beta-1.
\label{negativity}
\eeq
In our study of the nonclassicality of the photon-added squeezed kitten states we employ the negativity $\delta_{W}$ extensively.
\subsection{The Husimi $Q$-function}
\label{Husimi} 
The  Husimi $Q$-function [\cite{H1940}] is a normalized quasiprobability distribution in the phase space. Defined as the 
diagonal expectation value of the density matrix in an arbitrary coherent state 
\beq
Q(\beta,\beta^{*}) = \frac{1}{\pi} \bra\beta\rho \ket\beta, \quad \int Q(\beta, \beta^{*}) \mathrm{d}^{2}\beta = 1,
\label{Q_defn}
\eeq
it is a positive semi-definite smooth quantity that maintains 
 the following bounds: $0 \leq Q(\beta,\beta^{*}) \leq {\pi}^{-1}$.
The $Q$-function is also  obtained [\cite{SZ2001}] by performing a smoothing operation on the $P$-representation
via the Gaussian weight of unit variance:
\beq
Q(\beta,\beta^{*}) = \frac{1}{\pi}\int P(\gamma,\gamma^{*})  \exp(-|\beta-\gamma|^{2})  
\;\mathrm{d}^{2}\gamma.
\label{P_int_Q}
\eeq
Moreover, the $Q$-function 
may be regarded [\cite{SZ2001}] as the convolution of the $W$-distribution 
with a Gaussian kernel possessing a variance $1/2$ in the phase space:
\beq
Q(\beta,\beta^{*}) = \frac{2}{\pi}\int W(\gamma,\gamma^{*})  \exp(-2|\beta-\gamma|^{2})  
\;\mathrm{d}^{2}\gamma.
\label{W_int_Q}
\eeq

\par

For the pure state density matrix (\ref{density_matrix_mpa}) the  definition (\ref{Q_defn}) readily furnishes the $Q$-function 
on the phase space as
\beq
Q^{(\kappa)}(\beta,\beta^{*};t) =\dfrac{1}{\pi} \left(\mathcal{N}^{(\kappa)}\right)^{2} \exp(-|\beta|^{2})
\left | \sum_{n=\kappa}^{\infty} \!  \dfrac{\beta^{*n} }{\sqrt{n!}}(1+(-1)^{n-\kappa} \mathrm{c})\,
\mathsf{A}_{n,\kappa}(t) \right |^2,
\label{Q_function_pa}
\eeq 
which vanishes only at asymptotically large values of $|\beta|$.
Parallel to the case of the $W$-distribution the choice ($\mathrm{c}=\pm 1$) in (\ref{Q_function_pa}) 
 imparts a parity symmetric property of the $Q$-function as
\beq
Q^{(\kappa)}(\beta,\beta^{*};t)= Q^{(\kappa)}(-\beta,-\beta^{*};t)
\label{Q_symmetry}
\eeq
that underlies the observation of only the even number of kitten states in this instance.  In the  general case, however, both the even and odd number 
of kitten states are realized. 

\par
The polar phase density of the Husimi $Q$-function [\cite{TMG1993}] is a convenient tool towards 
describing the formation of transient kitten states as it conveys splitting of the density function into respective lobes. 
Defined as
\beq
Q(\widetilde{\theta}) = \int_{0}^{\infty} Q(\beta,\beta^{*}) |\beta|\; \mathrm{d}|\beta|, \quad \beta=|\beta| 
\exp(i\widetilde{\theta})
\label{Q_polar}
\eeq
the polar  density corresponding to the $Q$-function (\ref{Q_function_pa}) reads
\beq
Q^{(\kappa)}(\widetilde{\theta})=\dfrac{1}{2\pi}
\sum_{n,m=\kappa}^{\infty} \!  \dfrac{1 }{\sqrt{n!m!}} \left((m+n)/2\right)!\, \exp(-i (n-m) \widetilde{\theta}) \, \dm(t).
\label{Q_phase}
\eeq

\par

The Wehrl entropy [\cite{W1978}] provides a useful measure  
estimating the delocalization of the system in the phase space. Described via the $Q$ quasiprobability function
\beq
S_{Q}=-\int Q(\beta, \beta^{*}) \,\log Q(\beta, \beta^{*}) \, \mathrm{d}^{2}\beta
\label{WehrlDef}
\eeq  
it may be considered [\cite{BKK1995}] as a count of an equivalent number of widely separated coherent states necessary for covering the existing phase space occupation of the coupled oscillator. 
In the context of an initial  coherent state subject to the Kerr Hamiltonian (\ref{KerrH}) the unitary time 
evolution has been found [\cite{MTK1990}-\cite{MBWI2001}] to 
lead to the formations of the transient kitten states characterized by the 
superposition of a finite number of macroscopic coherent states. 
The presence of the anharmonic term in (\ref{KerrH}) leads to a periodicity of the Wehrl entropy 
$S_{Q}$ that develops a series of local minima at the rational submultiples 
of the said time period. The quantum superposition of the  
macroscopic  states are produced precisely at these local minima of the Wehrl entropy. 

\par

For our choice of the initial state (\ref{initial-state-mpa}) the photon-added Schr\"{o}dinger squeezed kitten 
configurations are also realized at the 
local minima of the Wehrl entropy $S_{Q}$ occurring  at the rational submultiples of the time period. Moreover, we observe that 
similar pattern of the photon-added squeezed kitten states  also emerge for the choice of the squeezed vacuum configuration
($\alpha = 0$) and a large squeezing parameter $r\sim 1.5$ in the initial state (\ref{initial-state-mpa}). We will describe 
it in the next subsection.

\subsection{Photon-added squeezed kitten states}
\label{kittenTom}
In the absence of decoherence the unitary time evolution of the pure state density matrix (\ref{density_matrix_mpa}) leads to a 
periodic behavior of its Wehrl entropy $S_{Q}$ (Figs. \ref{fig_kittens_mpa} $\mathsf{a}_{1}, \mathsf{a}_{2}, \mathsf{d}_{1}$), 
and other dynamical quantities such as negativity $\delta_{W}$ associated with the $W$-distribution 
(Figs. \ref{fig_kittens_mpa} $\mathsf{a}_{3}, \mathsf{a}_{4}$). One manifest characteristics in these diagrams is that a state, say,
with a larger squeezing parameter $r$ corresponds to higher values of the dynamical quantities $S_{Q}$ and $\delta_{W}$. This may be 
qualitatively understood as follows: Larger value of $r$ engenders increased occupation of higher Fourier modes of the coefficients 
(\ref{A_nk}), which, in turn, triggers the generation of more interference terms for the $W$-distribution (\ref{W_function_mpa}) 
as well as the $Q$-function (\ref{Q_function_pa}). This induces wider spread of the quasiprobability functions in the phase space, 
and, simultaneously, more numerous sign reversals of the $W$-distribution. Consequently, the Wehrl entropy and the negativity 
register increments with the enlargement of the squeezing parameter. Accretion of the number ($\kappa$) of photons added 
to the state produces the same qualitative effect. 

\par

These diagrams (Figs. \ref{fig_kittens_mpa} $\mathsf{a}_{1}-\mathsf{a}_{4}, \mathsf{d}_{1}$) display  a series of local
minima at rational submultiples of the time period. At these instants  the evolving state 
of the nonlinear Kerr oscillator coincides with the superposition of a finite number of photon-added squeezed coherent states. To 
demonstrate the above transitory convergence of states, we  construct superpositions of the photon-added squeezed kitten states and 
obtain their Hilbert-Schmidt distance measures [\cite{DMMW2000}] from the evolving Kerr oscillator states at the relevant time limits. 
The vanishing of the distances between the two sets of states  establishes their transient indistinguishability.  
The Hilbert-Schmidt distance between the states characterized by the density matrices $\rho_{1}$ and $\rho_{2}$ is defined as 
$\left(d_{\mathrm{HS}}(\rho_1, \rho_2)\right)^{2} \equiv \mathrm{Tr} \left( \rho_{1} -\rho_{2} \right)^{2}$, and  may be expressed
[\cite{DMMW2000}] via the corresponding $W$-distributions.

\par

The fiducial marker state occurring as a superposition of $\kappa$-photon added $p$ squeezed kittens, and the corresponding 
density matrix  may be listed as
\beq
\ket{\widetilde{\psi}^{(\kappa)}}  =
\widetilde{\mathcal{C}}^{(\kappa)} \sum_{j=0}^{p-1} f_{j} \,
a^{\dagger \kappa}\ket{\xi_{j},\alpha_{j}}, \quad
\widetilde{\rho}^{\,(\kappa)}= \ket{\widetilde{\psi}^{(\kappa)}} \bra{\widetilde{\psi}^{(\kappa)}},
\label{mpa_ref}
\eeq
where the complex coordinates describing the ensemble of kitten states read:
\beq
\xi_{j} =\xi \exp(2i(\widetilde{\vartheta}+2\pi j/p)), \quad \alpha_{j} 
=\alpha \exp(i(\widetilde{\vartheta}+2\pi j/p)),
\quad \nu_{j}= \nu \exp(2i(\widetilde{\vartheta}+2\pi j/p)).
\label{comp_co}
\eeq
\begin{figure}
\captionsetup[subfigure]{labelformat=empty}
\subfloat[(a$_{1}$)]{\includegraphics[width=4cm,height=2.8cm]{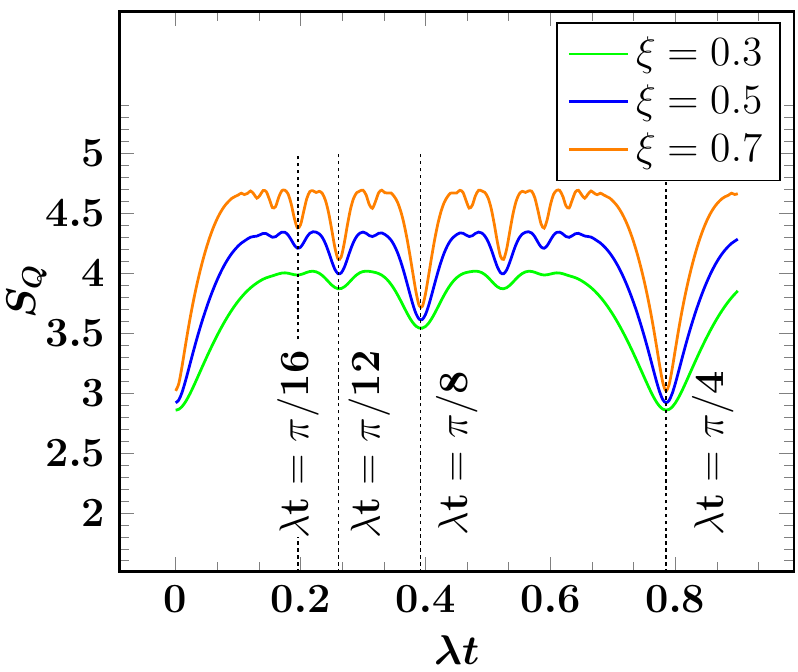}} 
\quad 
\captionsetup[subfigure]{labelformat=empty}
\subfloat[(a$_{2}$)]{\includegraphics[width=4cm,height=2.8cm]{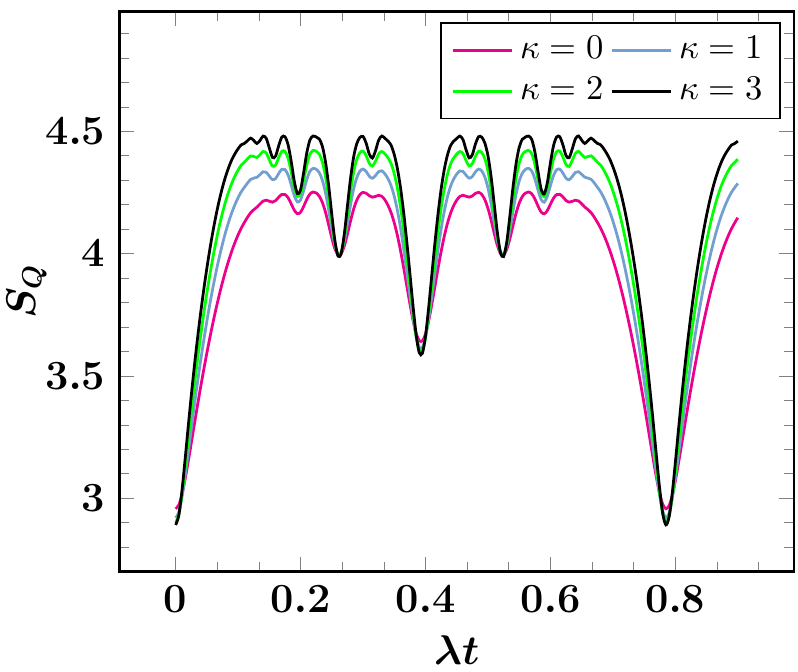}} 
\quad  
\captionsetup[subfigure]{labelformat=empty}
\subfloat[(a$_{3}$)]{\includegraphics[width=3.8cm,height=2.8cm]{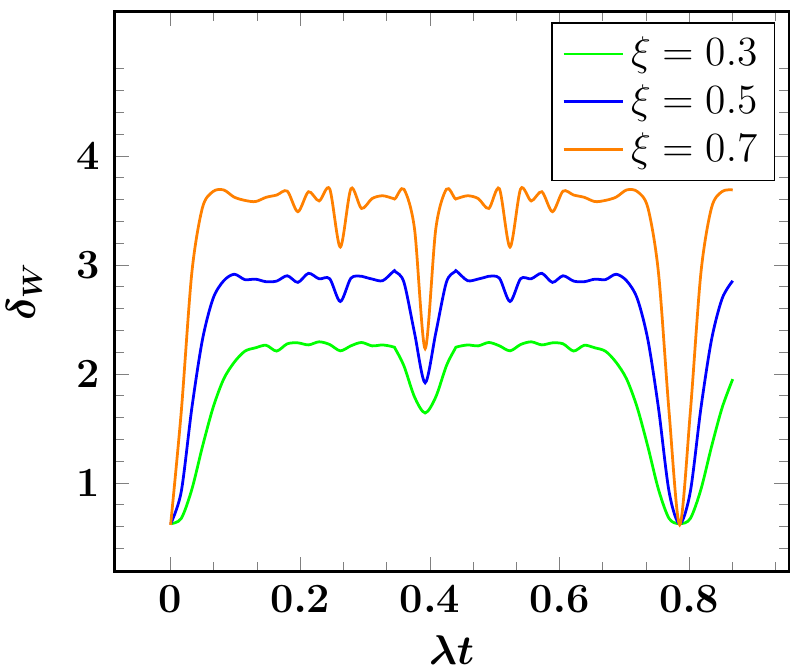}} 
\captionsetup[subfigure]{labelformat=empty}
\subfloat[(a$_{4}$)]{\includegraphics[width=4cm,height=2.8cm]{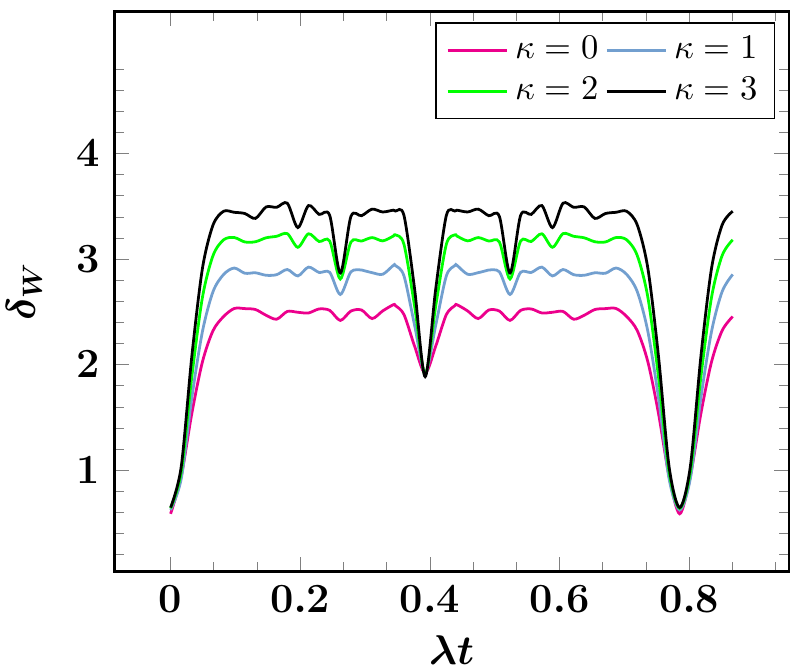}} 
\quad 
\\ \\ \\
\captionsetup[subfigure]{labelformat=empty}
\subfloat[(b$_{1}$)]{\includegraphics[scale=0.32]{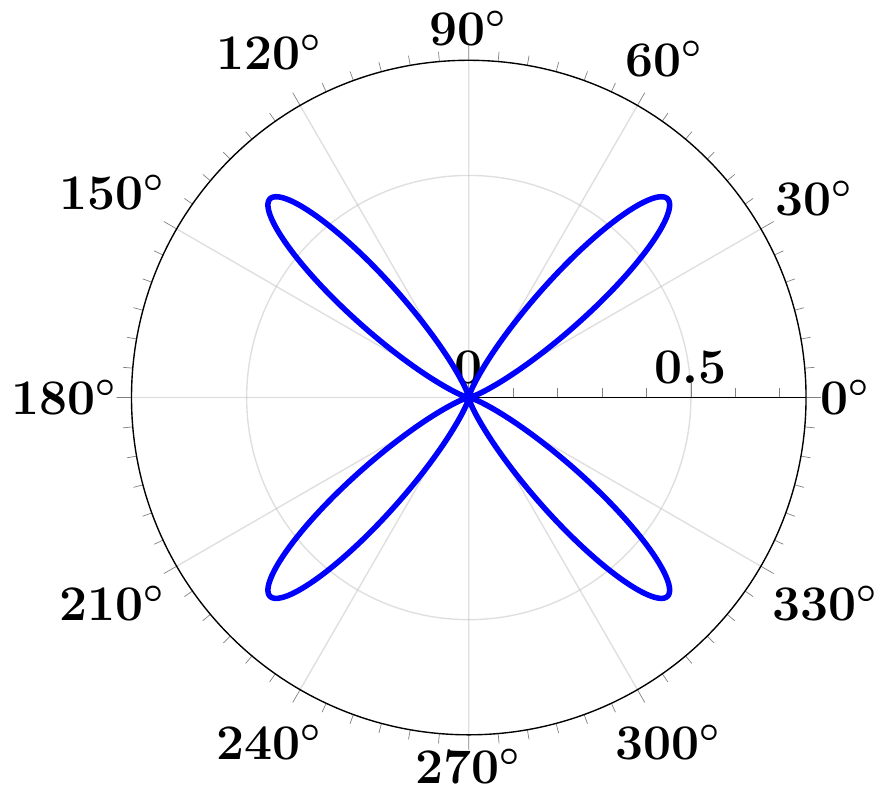}} 
\captionsetup[subfigure]{labelformat=empty}
\subfloat[(b$_{2}$)]{\includegraphics[scale=0.32]{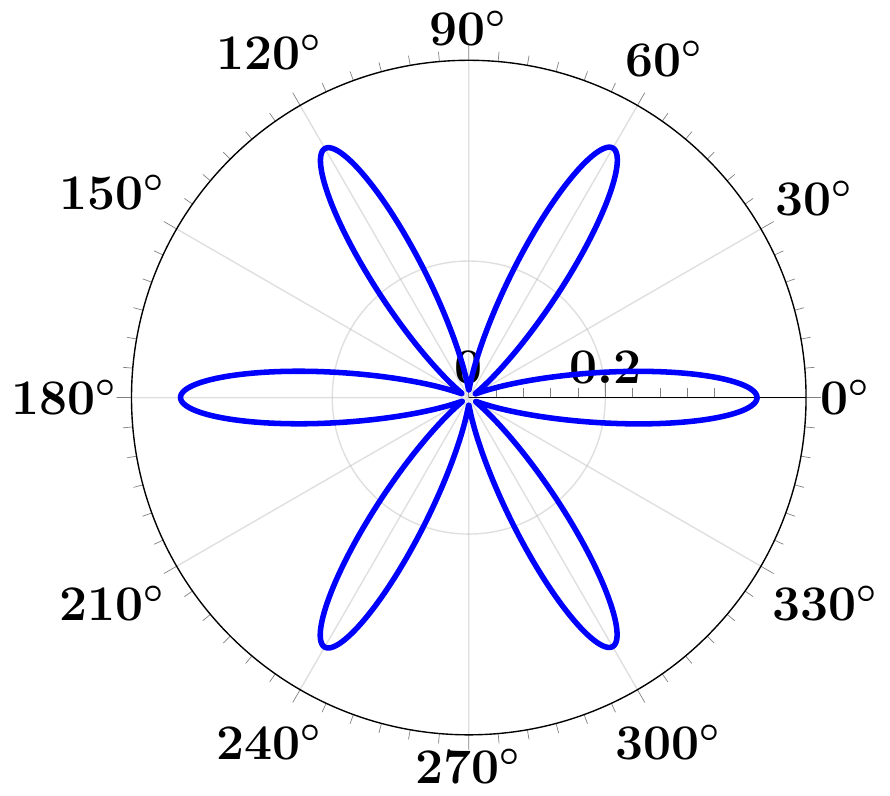}} 
\captionsetup[subfigure]{labelformat=empty}
\subfloat[(b$_{3}$)]{\includegraphics[scale=0.32]{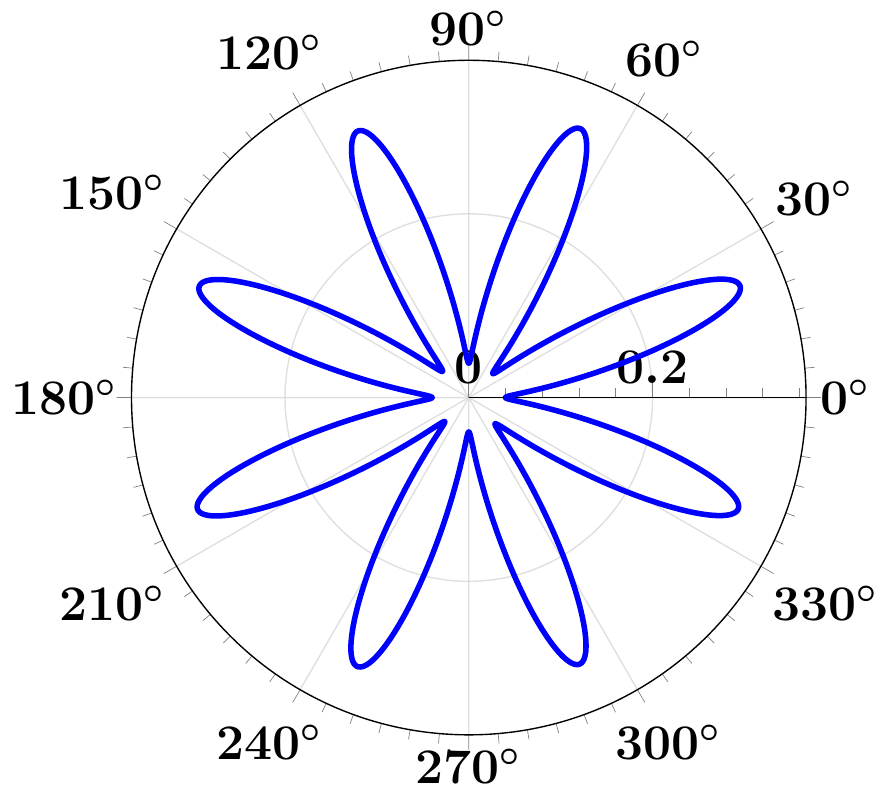}}
\captionsetup[subfigure]{labelformat=empty}
\subfloat[(c$_{1}$)]{\includegraphics[scale=0.29]{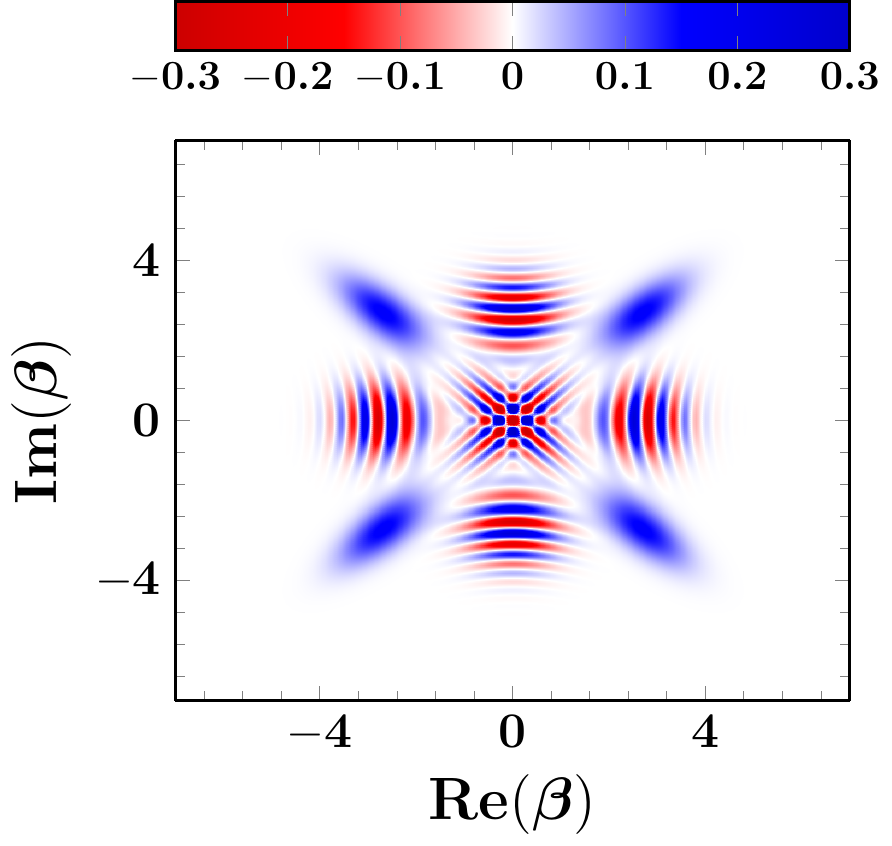}} 
\captionsetup[subfigure]{labelformat=empty}
\subfloat[(c$_{2}$)]{\includegraphics[scale=0.29]{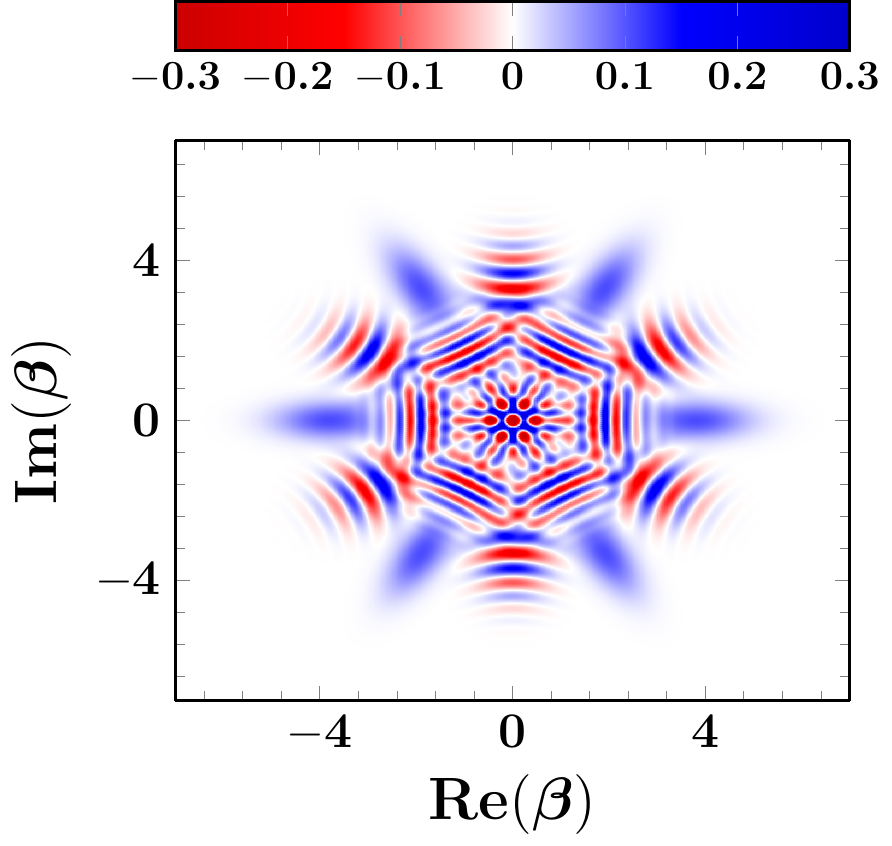}} 
\captionsetup[subfigure]{labelformat=empty}
\subfloat[(c$_{3}$)]{\includegraphics[scale=0.29]{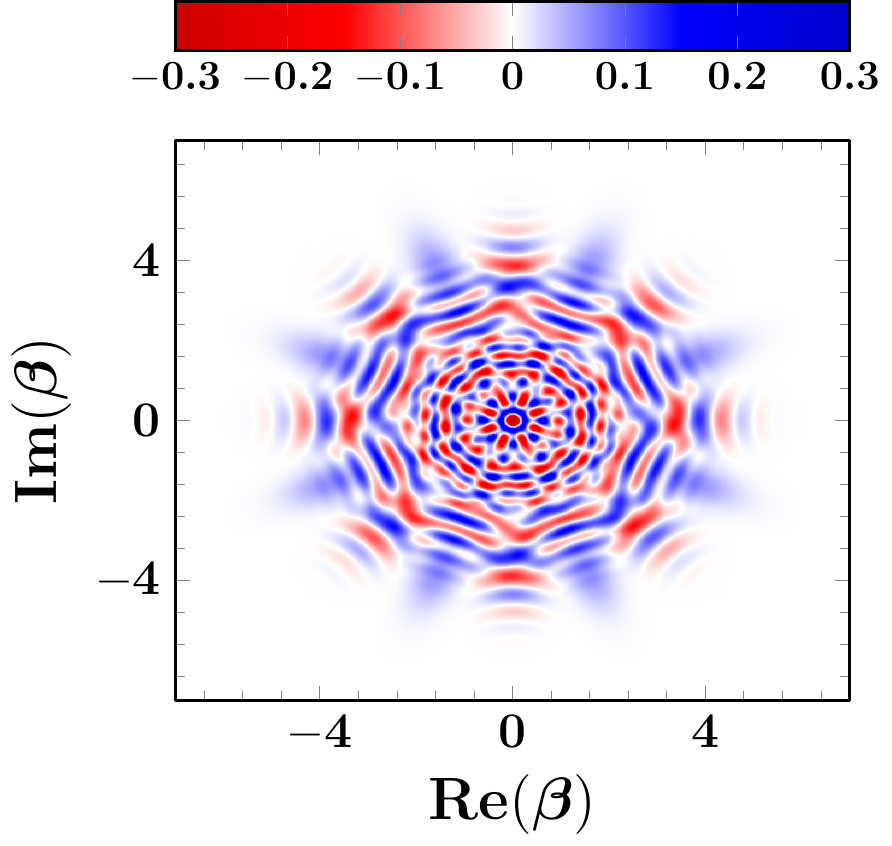}} \\ \\ \\
\captionsetup[subfigure]{labelformat=empty}
\subfloat[(d$_{1}$)]{\includegraphics[width=3.8cm,height=2.8cm]{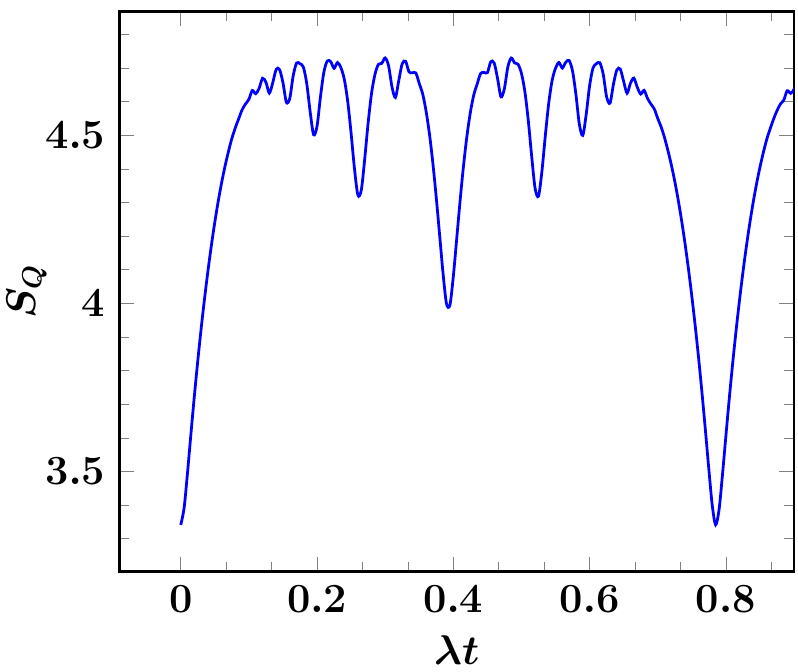}} \quad
\captionsetup[subfigure]{labelformat=empty}
\subfloat[(d$_{2}$)]{\includegraphics[scale=0.32]{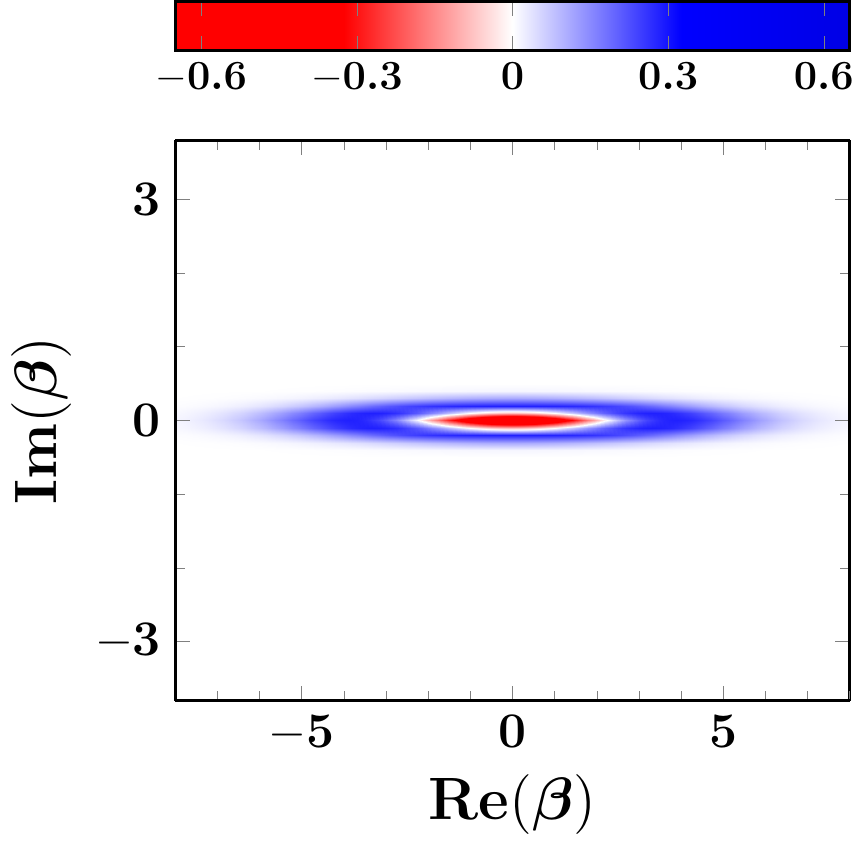}} \quad
\captionsetup[subfigure]{labelformat=empty}
\subfloat[(d$_{3}$)]{\includegraphics[scale=0.32]{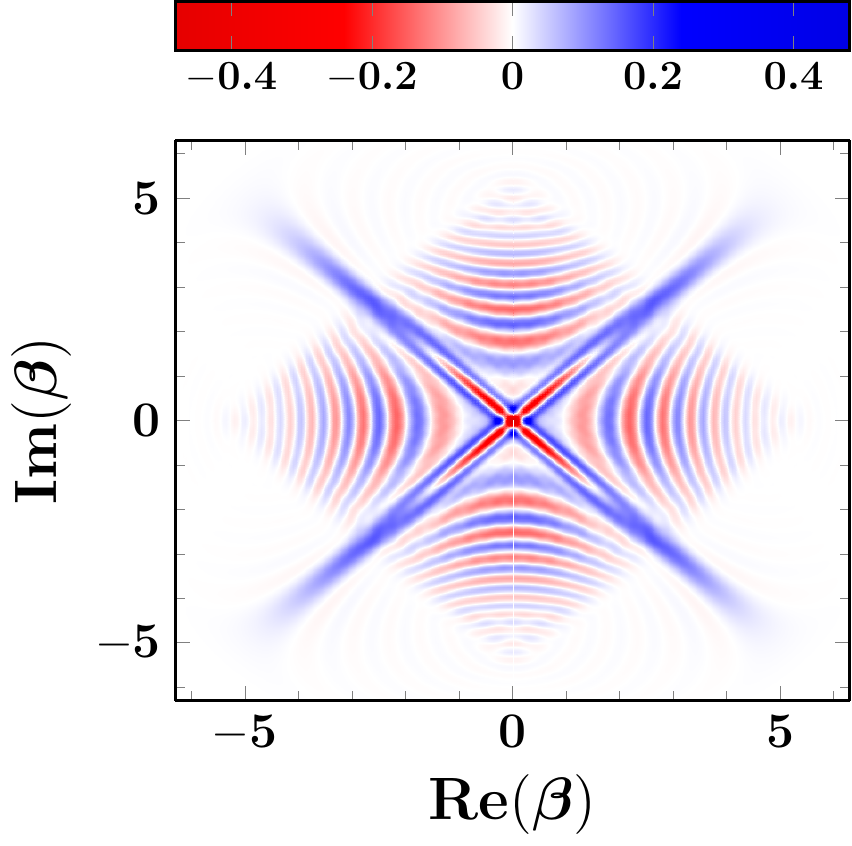}} \quad
\captionsetup[subfigure]{labelformat=empty}
\subfloat[(d$_{4}$)]{\includegraphics[scale=0.32]{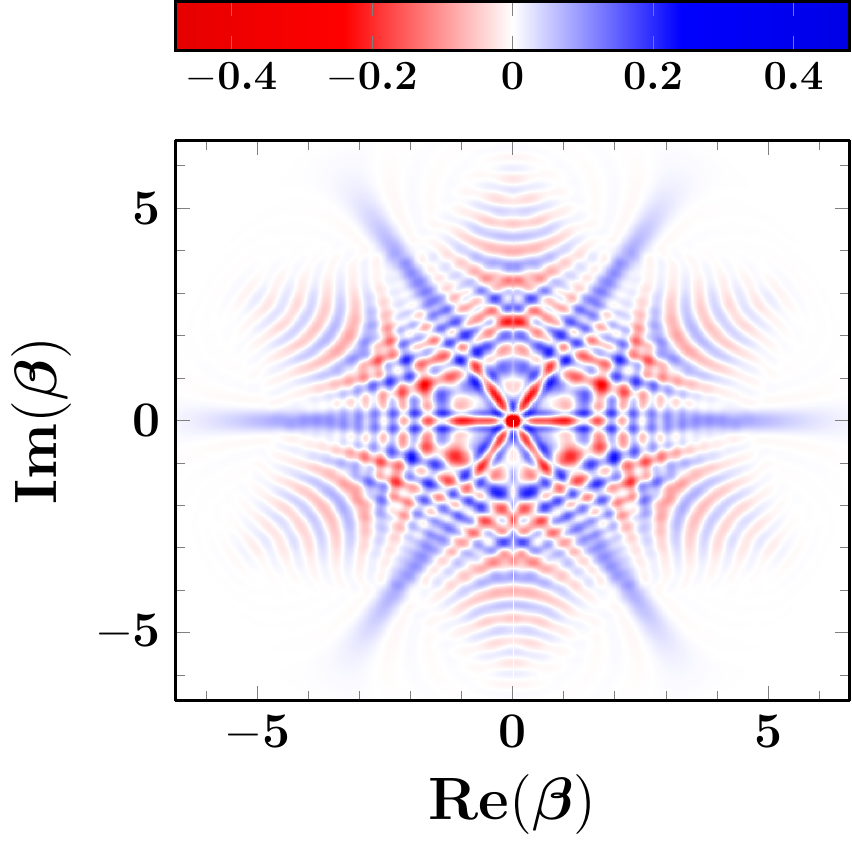}} \quad
\captionsetup[subfigure]{labelformat=empty}
\subfloat[(d$_{5}$)]{\includegraphics[scale=0.32]{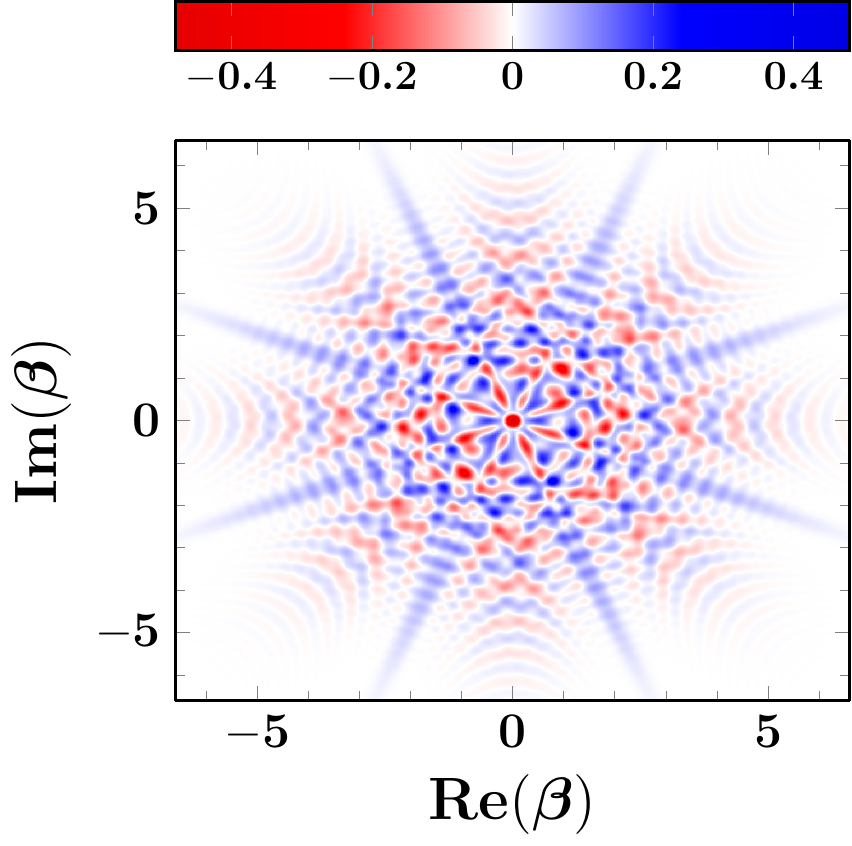}} 
\caption{For the selection of parameters $\mathrm{c} = 1, \delta = 1$ maintained throughout these diagrams, the plots 
($\mathsf{a}_{1}$) and ($\mathsf{a}_{2}$) 
refer, respectively,  to the Wehrl entropy  of the single photon-added case for $\alpha=2$ with varying squeezing parameter $\xi$, and that of the multiple photon-added cases for the variables $\alpha=2, \xi=0.5$. Retaining the choice of the parameters as in 
$(\mathsf{a}_{1})  ((\mathsf{a}_{2}))$, the diagram $(\mathsf{a}_{3}) ((\mathsf{a}_{4}))$ studies the negativity of the Wigner distribution.  Comparatively large values of each of the parameters  $\xi$ and $\kappa$ enhance the population of the higher modes as well as the spread of  the interference domain in the phase space. These effects, in turn, augment the observed values of $S_{Q}$ and $\delta_{W}$. The Figs. $(\mathsf{b}_{1}, \mathsf{b}_{2}, \mathsf{b}_{3}$) and $(\mathsf{c}_{1}, \mathsf{c}_{2}, \mathsf{c}_{3})$ correspond, respectively,
to the scaled times $(\frac{\pi}{8}, \frac{\pi}{12}, \frac{\pi}{16})$ 
showing the polar density $Q(\widetilde{\theta})$  and the $W$-distribution for the phase space variables 
$\alpha=2, \xi=0.5$. The formation of the transient kitten structures for the evolving photon-added squeezed coherent
states are evident therein. The  negativity $\delta_{W}$ corresponding to $(\mathsf{c}_{1}, \mathsf{c}_{2}, \mathsf{c}_{3})$ reads $(1.9224, 2.6665, 2.8487)$
depicting a systematic increase of nonclassicality with growing complexity of the quantum states that necessitates enlarged 
fluctuations of the higher Fourier modes of the density matrix. Another observation is that even for the photon-added squeezed 
vacuum initial state ($\alpha=0$) the evolution to the kitten structures is manifest for the dominant values of the squeezing 
parameter $\xi$. The initial state $\ket{\psi_{\alpha=0}^{(\kappa = 1)}(0)}$ with high squeezing parameter  $\xi=1.5$ develops
($\mathsf{d}_{1}$), during its evolution, a periodic structure of the  Wehrl entropy, while its  Wigner distribution  at $t=0$
($\left.\delta_{W}\right|_{t=0} = 0.4209$) is depicted in $(\mathsf{d}_{2})$. The subsequent diagrams
$(\mathsf{d}_{3}-\mathsf{d}_{5})$ obviously display its evolving 
transitory photon-added squeezed vacuum kitten states at the scaled times $(\tfrac{\pi}{8},\tfrac{\pi}{12},\tfrac{\pi}{16})$. The corresponding negativity variables $\delta_{W}$ are given by $(2.8336, 3.1297, 3.1981)$ reinforcing the observation that incremental 
complexity of the quantum states leads to higher nonclassicality.}
\label{fig_kittens_mpa}
\end{figure}
The evaluation of the normalization constant $\widetilde{\mathcal{C}}^{(\kappa)}
= \Big(\sum_{j,\ell=0}^{p-1} f_{j} f^{*}_{\ell} \sum_{n=\kappa}^{\infty} \mathcal{A}_{n,\kappa}( \xi_{j},\alpha_{j})
\mathcal{A}^{*}_{n,\kappa}( \xi_{\ell},\alpha_{\ell}) \Big)^{-1/2}$
is aided by the following bilinear sum obtained via the identity (\ref{Hermite_identity_mpa}):
\bea
\sum_{n=\kappa}^{\infty} \mathcal{A}_{n,\kappa}( \xi_{j},\alpha_{j})
\mathcal{A}^{*}_{n,\kappa}(\xi_{\ell},\alpha_{\ell}) \!\!\!\!\! &=& \!\!\!\!\! \kappa!\,
\dfrac{\mu}{\sqrt{{\cal U}_{j \ell}}}
\exp\left(- \dfrac{N_{j \ell}}{{\cal U}_{j \ell}} \right) \sum_{p=0}^{\kappa} \binom{\kappa}{p}
\left(\dfrac{-\nu_{j} \nu_{\ell}^{*}}{2 \, {\cal U}_{j \ell}} \right)^{p}
 \; \sum_{q=0}^{p} \dfrac{1}{q!(p-q)!}  \times\nn \\
&\times& \left( \dfrac{{\cal U}_{j \ell}}{\nu_{j}\nu^{*}_{\ell}} \right)^{\frac{q}{2}}
\mathrm{H}_{q}\left(\dfrac{-i\alpha_{j}}{\sqrt{2 \mu \nu_{j}}} \right) 
\mathrm{H}_{2p-q} \left( \dfrac{-i(\alpha_{j} \nu_{\ell}^{*} + \alpha_{\ell}^{*} \mu)}{\sqrt{2 \mu \nu_{\ell}^{*}\, 
	{\cal U}_{j \ell}}}\right),
\label{AjAl}
\eea
where the coefficients have the following structures:
\beq
{\cal U}_{j \ell} = \mu^{2}-\nu_{j} \nu_{\ell}^{*}, \quad N_{j\ell}=|\alpha|^{2}\, {\cal U}_{j \ell} 
- \alpha_{j}\alpha^{*}_{\ell}+\dfrac{1}{2\mu}
\left( {\cal U}_{j \ell}\,(\alpha_{j}^{2}\nu^{*}_{j}+\alpha_{\ell}^{*2}\nu_{\ell})
-\alpha_{j}^{2}\nu^{*}_{\ell}-\alpha_{\ell}^{*2}\nu_{j} \right).
\label{UNjl}
\eeq
The Wigner distribution of the state (\ref{mpa_ref}) is given by
\beq
\widetilde{W}^{\,(\kappa)}(\beta,\beta^{*}) = \left(\widetilde{\mathcal{C}}^{\,(\kappa)}\right)^{2}
\sum_{j,\ell=0}^{p-1} f_{j} f^{*}_{\ell} 
\widetilde{W}^{\,(\kappa)}_{{\mbox{\fontshape{it}\scriptsize{j\,$\ell$}}}}(\beta,\beta^{*}),
\label{W_kitten}
\eeq
where the $W$-distribution pertinent to the projection operator reads
\bea
\widetilde{W}^{\,(\kappa)}_{{\mbox{\fontshape{it}\scriptsize{j\,$\ell$}}}}(\beta,\beta^{*}) &=& \dfrac{2}{\pi} 
\sum_{n=0}^{\infty} (-1)^{n} \braket{\beta,n| a^{\dagger \kappa} |\xi_{j},\alpha_{j}}
\braket{\xi_{\ell},\alpha_{\ell}| a^{ \kappa} |\beta,n}\nn\\
&=& \dfrac{2}{\pi} \exp(-2|\beta|^{2})
\sum_{n,m=\kappa}^{\infty} \! \! \!  \dfrac{(2\beta^{*})^{n} (2\beta)^{m}}{\sqrt{n!m!}} 
{}_2F_0\Big( -n,-m;\phantom{}_{-} ; -\dfrac{1}{4|\beta|^{2}}\Big) \quad \times \nn\\ 
&& \times \;\;\mathcal{A}_{n,\kappa}( \xi_{j},\alpha_{j})
\mathcal{A}^{*}_{m,\kappa}( \xi_{\ell},\alpha_{\ell}).
\label{W_jlProj}
\eea
We now provide an explicit evaluation of the infinite sum in (\ref{W_jlProj}), say for the $\kappa = 1$ case, and 
determine the corresponding $W$-distribution:
\beq
\widetilde{W}^{\,(\kappa =1)}_{{\mbox{\fontshape{it}\scriptsize{j\,$\ell$}}}}(\beta,\beta^{*}) =
 \dfrac{2\, H_{j\ell}}{\pi \, ({\cal U}_{j \ell} )^{\frac{5}{2}} } \,
 \exp \left(-2|\beta|^{2} - F_{j\ell}\right).
 \label{W_initial_time_spa}
\eeq
The  exponent and the amplitude factors in (\ref{W_initial_time_spa}) are of the following form:
\bea
 F_{j\ell} \!\!\! &=& \!\!\! |\alpha|^{2} + \dfrac{1}{2\mu} (\alpha_{j}^{2}\nu^{*}_{j} + \alpha_{\ell}^{*2}\nu_{\ell}) -
 \dfrac{2\beta^{*}}{\mu} \mathsf{F}_{j} - \dfrac{2\beta}{\mu} \mathsf{F}_{\ell}^{*} + \dfrac{1}{{\cal U}_{j \ell}} 
 \left\lgroup \mathcal{F}_{j} \mathcal{F}_{\ell}^{*}
 -\dfrac{\nu^{*}_{\ell} \mathcal{F}_{j}^{2}}{2\mu}
  -\dfrac{\nu_{j}\mathcal{F}_{\ell}^{*2}}{2\mu} \right\rgroup,\nn \\
H_{j\ell} \!\!\! &=& \!\!\! \left(\mu^{2}+ \nu_{j} \nu^{*}_{\ell} \right) 
\mathcal{F}_{j} \mathcal{F}_{\ell}^{*}
- \mu \nu^{*}_{\ell} \mathcal{F}_{j}^{2}
-\mu \nu_{j} \mathcal{F}_{\ell}^{*2} - \, {\cal U}_{j \ell} 
\left\lgroup \alpha_{j} \alpha^{*}_{\ell} 
- \mu^{2}(4|\beta|^{2}-1) \right. \nn \\ 
&& \left. -\mathcal{F}_{j} \mathcal{F}_{\ell}^{*}
+ 2\mu \beta^{*} \mathcal{F}_{j}  + 2\mu \beta \mathcal{F}_{\ell}^{*} \right\rgroup,
\label{F_G} 
\eea
where the phase space variables read: 
$\mathsf{F}_{j} = \alpha_{j} +  \beta^{*} \nu_{j} $, $ \mathcal{F}_{j} = \alpha_{j} + 2 \beta^{*} \nu_{j}$. 
The normalization constant may be evaluated as follows:
\bea
\widetilde{\mathcal{C}}^{\,(\kappa =1)} = \lvast \lgroup \sum_{j,\ell =0}^{p-1} 
\dfrac{f_{j} f^{*}_{\ell}}{({\cal U}_{j \ell})^{\frac{5}{2}}} 
\exp\left(- \dfrac{N_{j\ell}}{{\cal U}_{j \ell}} 
\right) \svast \lgroup {\cal U}_{j \ell}^{2} +  
\left(\alpha_{j} \alpha^{*}_{\ell} (\mu^{2}+\nu_{j} \nu^{*}_{\ell})
 +\mu(\alpha_{j}^{2}\nu^{*}_{\ell} + \alpha_{\ell}^{*2}\nu_{j})
+   \nu_{j}\nu^{*}_{\ell} \, {\cal U}_{j \ell} \right) \svast \rgroup \lvast \rgroup
 ^{-\frac{1}{2}} \!\!.
 \label{Wjl_norm}
 \eea
  The quasiprobability distribution for the benchmark state $\widetilde{W}^{\,(\kappa = 1)}(\beta,\beta^{*})$ for the single 
 photon-added case now follows from (\ref{W_kitten}, \ref{W_initial_time_spa}, \ref{Wjl_norm}).
In the derivation of the  expression (\ref{W_initial_time_spa}) we have enlisted the following identities [\cite{AAR1999}]: 
 \bea
 \sum_{n=0}^{\infty} \dfrac{(n+k+1)\mathsf{t}^{n}}{n!} \mathrm{H}_{n+k}(\mathsf{x}) &=& 
 \left( \mathsf{t}\, \mathrm{H}_{k+1}(\mathsf{x}-\mathsf{t}) 
 + (k+1) \mathrm{H}_{k}(\mathsf{x}-\mathsf{t})\right) \exp(2 \mathsf{t} \mathsf{x}-\mathsf{t}^{2}), 
 \label{Hermite_W_id_1}\\
\sum_{n=0}^{\infty}\dfrac{(n+1) \mathsf{t}^{n}}{2^{n} n!}  \mathrm{H}_{n}(\mathsf{x}) \mathrm{H}_{n}(\mathsf{y}) &=& 
  \dfrac{1+ 2 \,\mathsf{t}(1+\mathsf{t}^{2})\,\mathsf{x y}-\mathsf{t}^{2}\left(1 + 2 \,(\mathsf{x}^{2} + \mathsf{y}^{2})\right)}{(1-\mathsf{t}^{2})^{5/2}} 
  \; \mathcal{E}_{\mathsf{t}}(\mathsf{x},\mathsf{y}),
  \label{Hermite_W_id_2}
\\
 \sum_{n=0}^{\infty}\dfrac{\mathsf{t}^{n}}{2^{n}n!} \mathrm{H}_{n+k}(\mathsf{x}) \mathrm{H}_{n}(\mathsf{y}) &=& 
 \dfrac{1}{(1-\mathsf{t}^{2})^{\tfrac{k+1}{2}}}\; 
 \,\mathrm{H}_{k}\left(\dfrac{\mathsf{x}-\mathsf{t y}}{\sqrt{1-\mathsf{t}^{2}}}\right)
  \mathcal{E}_{\mathsf{t}}(\mathsf{x},\mathsf{y}), 
 \label{Hermite_W_id_3} \\
   \sum_{n=0}^{\infty}\dfrac{\mathsf{t}^{n}}{2^{n}n!} 
\mathrm{H}_{n+\kappa}(\mathsf{x}) \mathrm{H}_{n+\kappa}(\mathsf{y}) &=& 
\dfrac{(-\mathsf{t})^\kappa}
{(1-\mathsf{t}^{2})^{\tfrac{2\kappa+1}{2}}} \,
\mathcal{E}_{\mathsf{t}}(\mathsf{x},\mathsf{y}) 
 \sum_{p=0}^{\kappa} \binom{\kappa}{p} \left( \dfrac{\sqrt{1-\mathsf{t}^{2}}}{\mathsf{t}}\right)^{p}
\mathrm{H}_{p}(\mathsf{x}) \, \mathrm{H}_{2\kappa-p}\left(\dfrac{\mathsf{x}\mathsf{t}-\mathsf{y}}{\sqrt{1-\mathsf{t}^{2}}}\right). \qquad
  \label{Hermite_W_id_4} 
 \eea

\par

The Hilbert-Schmidt distance [\cite{DMMW2000}] between the evolving state (\ref{W_function_mpa}) of the Kerr oscillator  
at the specified times and the fiducial state (\ref{W_kitten}) comprising of the superposition of  the photon-added squeezed 
coherent states may now be expressed as follows:
\beq
\left(d_{\mathrm{HS}}\left(\rho^{(\kappa)}(t), \widetilde{\rho}^{\,(\kappa)}\right)\right)^{2}  
= \pi \int 
\left(W^{(\kappa)}(\beta,\beta^{*};t)-\widetilde{W}^{\,(\kappa)}(\beta,\beta^{*})\right)^2 \;\mathrm{d}^2 \beta.
\label{HS_distance}
\eeq
When the convergence of the two states is realized at the rational submultiples of the time period of the Wehrl entropy $S_{Q}$
of the Kerr oscillator (Figs. \ref{fig_kittens_mpa} $\mathsf{a}_{1}, \mathsf{a}_{2}$), the distance measure (\ref{HS_distance}) 
goes to the precise zero limit. 
In specific, we now consider the case $\mathrm{c} =1$, where only the even number of kittens ($p=2,4,6...$) are produced at 
times $\lambda t = \frac{\pi}{2p}$. This can be observed from the evolution of the polar density $Q(\widetilde{\theta})$ as well 
as the $W$-distribution given in Figs. \ref{fig_kittens_mpa} 
$(\mathsf{b_{1}, b_{2}, b_{3}})$ and $(\mathsf{c_{1}, c_{2}, c_{3}})$, respectively. 
These transitory states are identified with
the fiducial marker state (\ref{mpa_ref}) with appropriate coefficients that render the
 Hilbert-Schmidt distance measure (\ref{HS_distance}) vanish at the relevant time limits. For the  $\mathrm{c} =1$
case these coefficients are given as
\beq 
f_{j}= \exp\left( \dfrac{i\pi j^{2}(p+2)}{p} \right), \quad
\widetilde{\vartheta}=\dfrac{\pi}{2p}\left( p - 2\kappa +1 - \delta \right), \quad \delta = \dfrac{\omega}{\lambda},
\label{f-theta-kitten}
\eeq
where the supplementary angle of rotation $\widetilde{\vartheta}$  implements complete phase space overlap between the density 
matrix of the reference state (\ref{mpa_ref}) and that of the transient squeezed kitten states (\ref{density_matrix_mpa}) observed when the locally minimum Wehrl entropy configurations  are realized. At the submultiples of the time period, we, therefore, find a transient  
production of the Yurke-Stoler  type of states [\cite{YS1986}] which are a linear combination of equally phase-shifted photon-added
squeezed coherent states. 

\par

The appearances of the similar kitten-like patterns in the phase space are also noticed (Figs. \ref{fig_kittens_mpa} 
$\mathsf{d}_{1}-\mathsf{d}_{5}$) for the time evolution of the photon-added highly squeezed vacuum states $\alpha = 0, \xi \sim 1.5$. This result points towards another possibility. The well-known one mode Schwinger representation of the $su(1,1)$ algebra [\cite{SZ2001}] reads 
$[K_{0}, K_{\pm}] = \pm K_{\pm}, [K_{+}, K_{-}] = - 2 K_{0}, K_{+} \equiv \frac{1}{2} a^{\dagger\, 2}, K_{-} \equiv \frac{1}{2} a^{2},  K_{0} \equiv \frac{1}{4} (a a^{\dagger} + a^{\dagger} a)$, and, consequently, the squeezed vacuum states may be understood as particular realizations of coherent states of $su(1, 1)$ algebra [\cite{P1986}]. The present result, therefore, indicates the possibility of similar 
kitten-type formations existing in the evolution of other systems governed by the $su(1, 1)$ symmetry. This issue will be examined elsewhere.

\par
 
A general feature observed in the kitten-like organizations in the phase space is that higher complexity
necessitates more intensified occupation of higher oscillatory Fourier modes, say, in the $W$-distribution. This, in turn, triggers 
an increased value of the negativity $\delta_{W}$  signifying the presence of more nonclassicality of the state.   
Maintaining the parametric choice $\mathrm{c} = 1, \kappa = 1$, we plot in Fig. \ref{HS-0} the time variation of the Hilbert-Schmidt distance between the evolving state (\ref{density_matrix_mpa}) and the fiducial state (\ref{mpa_ref}) for, say $p = 4$  squeezed 
kitten formations. We note that as the displacement parameter $\alpha$ increases, the distance  collapses  to the null value
$d_{\mathrm{HS}}\left(\rho^{(\kappa)}(t), \widetilde{\rho}^{\,(\kappa)}\right) \rightarrow 0$ with \textit{growing sharpness} at the 
appropriate submultiple  of the time period: $\lambda t = \frac{\pi}{8}$. As enlarged value of $\alpha$ indicates more macroscopic 
nature of the state, the rising rapidity of the collapse for higher $\alpha$ points towards extremely short-lived nature of the 
kitten states for the macroscopic systems.
\begin{figure}[H]
	\captionsetup[subfigure]{labelformat=empty}
	\subfloat[(a$_{1}$)]{\includegraphics[width=4.5cm,height=3.5cm]
		{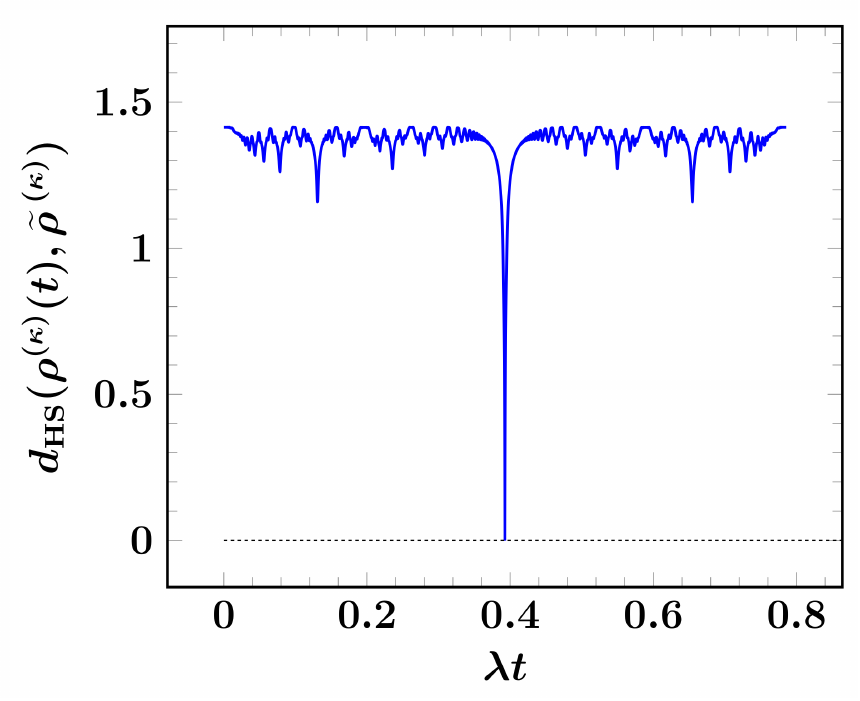}} \hspace{0.6cm}
		\captionsetup[subfigure]{labelformat=empty}
	\subfloat[(a$_{2}$)]{\includegraphics[width=4.5cm,height=3.5cm]
	{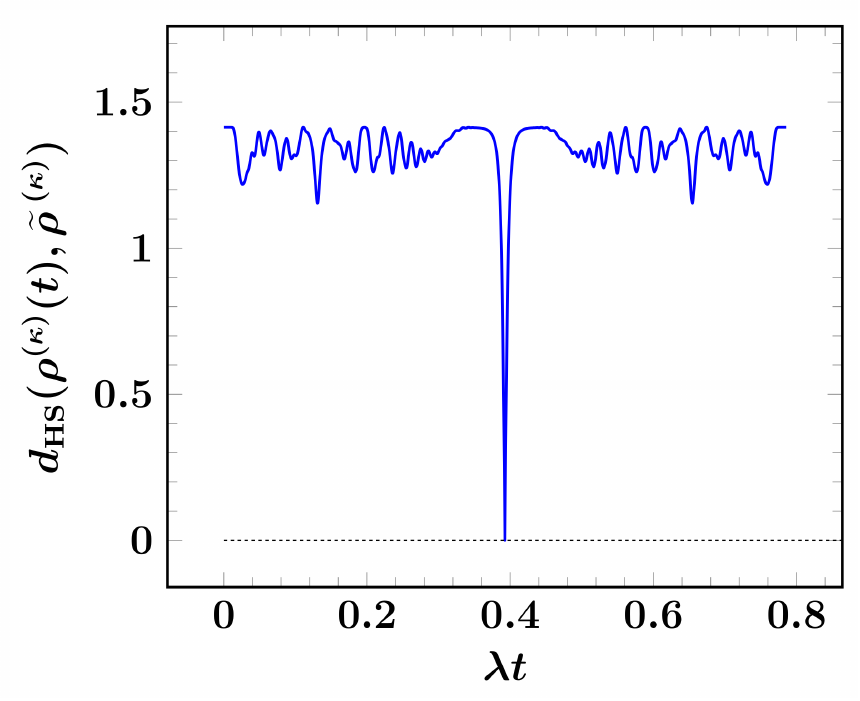}} \hspace{0.6cm}
	\captionsetup[subfigure]{labelformat=empty}
	\subfloat[(a$_{3}$)]{\includegraphics[width=4.5cm,height=3.5cm]
		{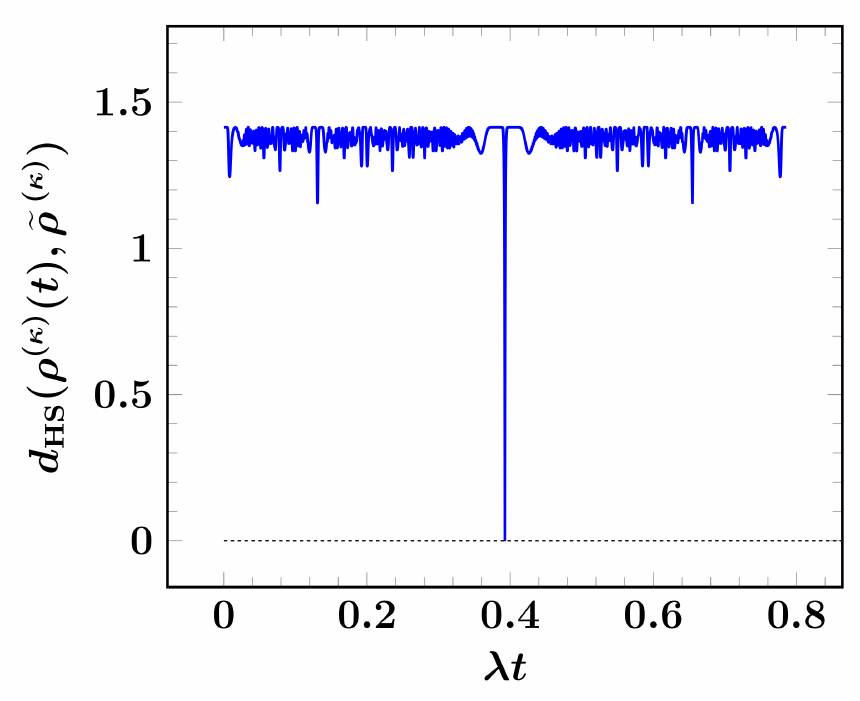}}
	\caption{ 
	The evolution of the Hilbert-Schmidt distance $d_{\mathrm{HS}}\left(\rho^{(\kappa)}(t), \widetilde{\rho}^{\,(\kappa)}\right)$ for 
	$\mathrm{c} = 1, \delta = 1,  \kappa = 1$ case
	is studied for $p =4$ kitten formations realized at time $\lambda t = \frac{\pi}{8}$. The diagram ($\mathsf{a}_{1}$) corresponds to 
	the phase space variables ($\alpha=0, \xi = 1.5$). The plots
		($\mathsf{a}_{2}$) and ($\mathsf{a}_{3}$) refer to $\xi=0.5$ while maintaining the displacement parameter to $\alpha=2$ and 
		$\alpha=4$, respectively.}
		\label{HS-0}
\end{figure}
\subsection{Optical tomogram}
\label{OT}
\begin{figure}[H]
	\captionsetup[subfigure]{labelformat=empty}
	\subfloat[(a$_{1}$)]{\includegraphics[scale=0.4]{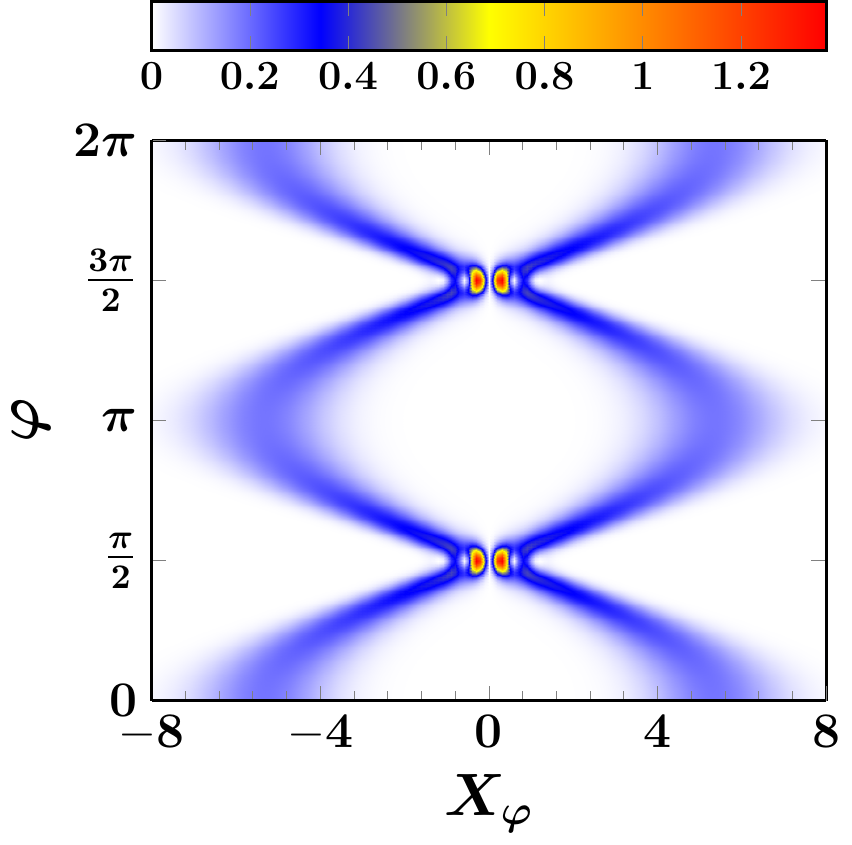}} \; \;
	\captionsetup[subfigure]{labelformat=empty}
	\subfloat[(a$_{2}$)]{\includegraphics[scale=0.4]{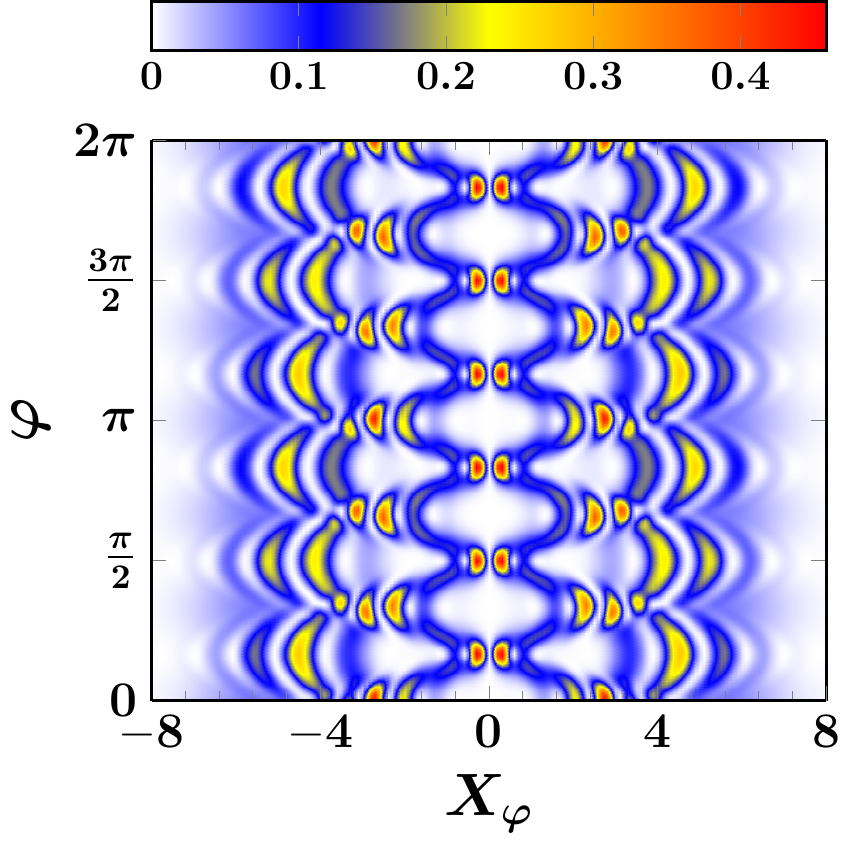}} \; \;
	\captionsetup[subfigure]{labelformat=empty}
	\subfloat[(a$_{3}$)]{\includegraphics[scale=0.4]{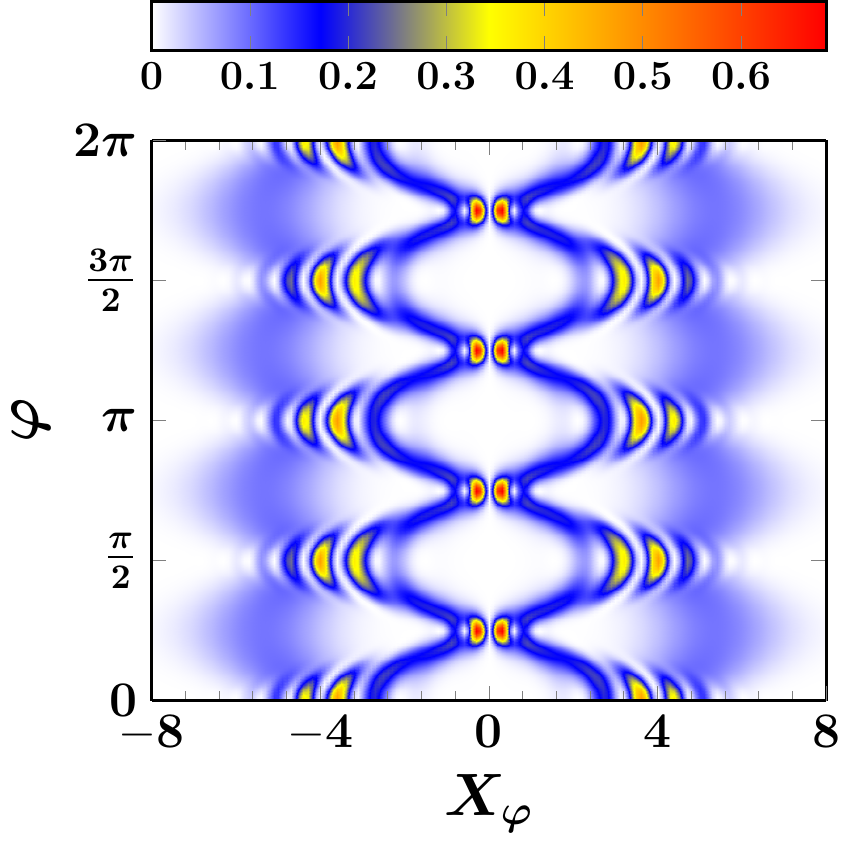}} \; \;
	\captionsetup[subfigure]{labelformat=empty}
	\subfloat[(a$_{4}$)]{\includegraphics[scale=0.4]{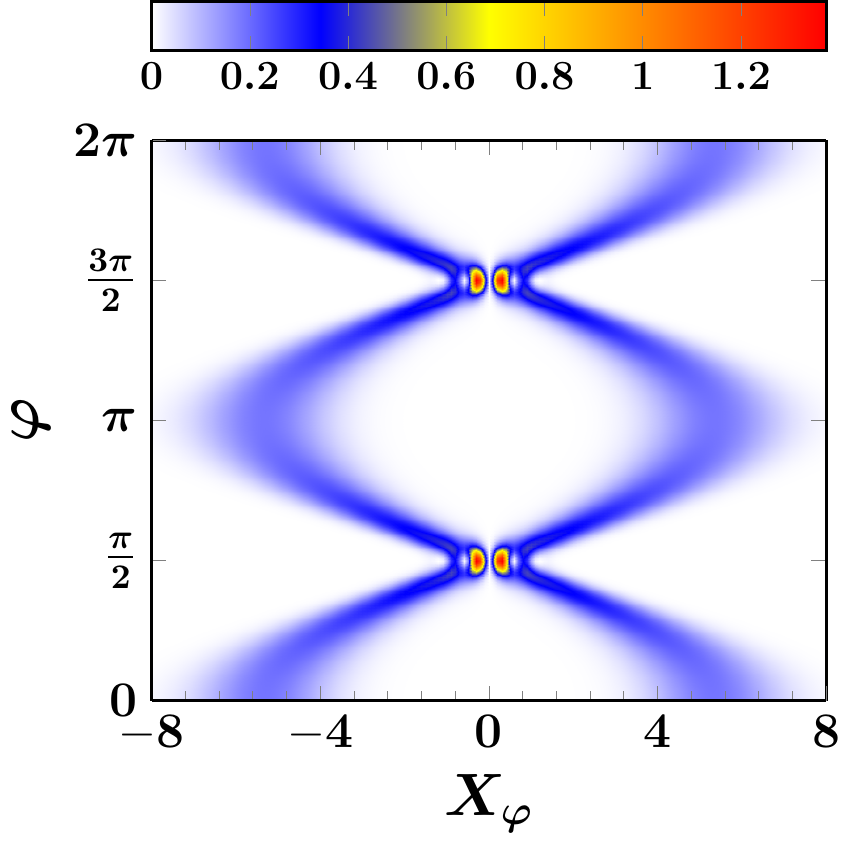}}
	\caption{The diagrams $(\mathsf{a}_{1}, \mathsf{a}_{2}, \mathsf{a}_{3}, \mathsf{a}_{4})$ plot the optical tomograms for the instance
	$\mathrm{c} =1, \kappa=1, \delta = 1$ corresponding to the scaled time $\lambda t$ at 
	$(0,\tfrac{\pi}{12},\tfrac{\pi}{8},\tfrac{\pi}{4})$, 
	respectively. The phase space variables read  $\alpha=2, \xi=0.5$.}
	\label{fig_OT_mpa}
\end{figure}

The  optical tomography has been recently advanced as an important procedure towards measuring and reconstructing the quantum state 
of optical fields [\cite{VR1989}-\cite{LR2009}]. It is based on the one-to-one correspondence between the quasiprobability 
distributions and the relevant probability distribution of the arbitrarily rotated quadrature  of the field variable [\cite{VR1989}]. The optical tomogram contains all informations on the quantum system and provides its alternate description vis-\`{a}-vis the conventional density 
matrix representation [\cite{VR1989}, \cite{MMT1996}]. Experimentally, a series of measurements of the rotated quadrature variable is performed on an ensemble of identically prepared systems. The  histogram obtained therefrom facilitates reconstruction [\cite{SBRF1993}] of the quasiprobability distributions.

\par

The arbitrarily rotated quadrature variable and its eigenstate are defined as
\beq
\widehat{X}=\frac{1}{\sqrt{2}} \left(a \exp(-i\varphi)+ a^{\dagger} \exp(i\varphi)\right),
\qquad
\widehat{X} \ket{X,\varphi} = X \ket{X,\varphi}.
\label{angle_hermitian_operator} 
\eeq
The construction [\cite{BR1997}] of the eigenstate (\ref{angle_hermitian_operator})  of the rotated position operator reads
\beq
\ket{X,\varphi} = \dfrac{1}{\pi^{1/4}} 
\exp \left( -\dfrac{X^{2}}{2} - \dfrac{e^{i2\varphi}}{2} a^{\dagger ^{2}}
+ \sqrt{2} e^{i\varphi} X a^{\dagger}\right) \ket{0}.
\label{X_eigenstate}
\eeq
The optical tomogram $\Omega(X,\varphi)$ is defined [\cite{VR1989}] as the probability distribution 
of the rotated  quadrature operator $\widehat{X}$ expressed via the expectation value of the density matrix in the basis of 
the eigenstates (\ref{X_eigenstate}):
\beq
\Omega(X,\varphi) \equiv \braket{X,\varphi|\rho|X,\varphi}, \qquad
\int^{\infty}_{- \infty} \mathrm{d}X\, \Omega(X,\varphi)=1.
\label{optical_tomogram}
\eeq
It has been observed [\cite{RS2015}] that the optical tomogram $\Omega(X,\varphi)$  embodies signatures of 
revival and fractional revivals of a quantum system, and, therefore, may be suitably applied towards the study of the
generation of kitten states. The optical tomogram for our model reads:
\beq
\Omega^{(\kappa)}(X,\varphi; t)= \left(\mathcal{N}^{(\kappa)}\right)^{2}\; \dfrac{\exp(-X^{2})}{\sqrt{\pi}}
\Big| \sum_{n=\kappa}^{\infty} \dfrac{(1+(-1)^{n-\kappa}\mathrm{c})}{\sqrt{2^{n} \,n!}} \, 
\mathsf{A}_{n,\kappa}(t) \,\exp(-i n \varphi) \,
\mathrm{H}_{n}(X) \Big|^{2},
\label{OTt}
\eeq
where the normalization property (\ref{optical_tomogram}) follows from (\ref{norm-mpa}, \ref{state-t}), and the 
orthonormality of the Hermite polynomials: $\int \mathrm{d}X \exp(-X^{2}) \mathrm{H}_{n}(X) \mathrm{H}_{m}(X)
= \delta _{n, m} 2^{n}\sqrt{\pi}\, n!$. The rhs in (\ref{OTt}) establishes
the parity symmetric behavior $\Omega^{(\kappa)}(- X,\varphi; t) = \Omega^{(\kappa)}(X,\varphi; t)$ 
for the choice $\mathrm{c} = \pm 1$. The optical tomogram at $t=0$ may now 
be directly constructed as
\beq
\Omega(X,\varphi;t=0)= 
(\mathcal{N}^{(\kappa)})^{2} \,\dfrac{\exp(-X^{2})}{\sqrt{\pi}\mu 2^{\kappa}} 
\exp \left(-|\alpha|^{2}-\frac{\alpha^{2}\nu^{*}}{2\mu}-\frac{\alpha^{*2}\nu}{2\mu} \right) 
\lvert f_{\varphi}(\xi,\alpha)+ \mathrm{c} f_{\varphi}(\xi,-\alpha)\rvert^{2},
\label{Omega_t0}
\eeq
where the structure function reads
\bea
 f_{\varphi}(\xi,\alpha) = \dfrac{1}{\left( 1+ \dfrac{\nu}{\mu} \exp(-2i \varphi) \right)^{\frac{\kappa+1}{2}}} \;
  \exp\left(\!\! \dfrac{\nu X^{2}-\frac{\alpha^{2}}{2\mu}+\sqrt{2} \alpha X 
	\exp(i \varphi)}{\mu \exp(2i \varphi) + \nu } \!\!\right) 
\mathrm{H}_{\kappa} \!\!\left( \dfrac{X-\frac{\alpha }{\sqrt{2}\mu} 
	\exp(-i\varphi)}{\sqrt{1+ \dfrac{\nu}{\mu} \exp(-2i \varphi)}}\right). 
\label{f_t0}	
\eea
To obtain the above expression we enlisted the identity (\ref{Hermite_W_id_3}). Each structure function in (\ref{Omega_t0}) produces a filament-like formation in the optical tomogram given in 
Fig. \ref{fig_OT_mpa} $(\mathsf{a}_{1})$. In general, the squared modulus of each of the structure functions at instants of time equal to the submultiples of the period of the Wehrl entropy $S_{Q}$ becomes manifest in the optical tomogram 
$\Omega^{(\kappa)}(X,\varphi; t)$  as the corresponding filament. The ripples observed in Fig. \ref{fig_OT_mpa} $(\mathsf{a}_{1})$ may be ascribed to the interference pattern produced by the inner product of two structure functions in (\ref{Omega_t0}).

\par

Towards obtaining the tomogram at the scaled time $\lambda t=\pi/4$ we use the following identity:
\beq
\exp\left( -i \dfrac{\pi}{4} n^{2} \right) = \dfrac{1+(-1)^n}{2} i^n + \dfrac{1-(-1)^n}{2} 
\exp\left( -i \dfrac{\pi}{4} \right),
\label{n_linear_pi_four}
\eeq
which facilitates expressing the infinite sum (\ref{OTt}) as a closed form expression involving a finite number of structure functions as follows:
\beq
\Omega\Big(X,\varphi; \lambda t=\frac{\pi}{4}\Big)= 
(\mathcal{N}^{(\kappa)})^{2}\; \dfrac{\exp(-X^{2})}{\sqrt{\pi}\mu 2^{\kappa}}\; 
\exp\left(-|\alpha|^{2}-\frac{\alpha^{2}\nu^{*}}{2\mu}-\frac{\alpha^{*2}\nu}{2\mu} \right) |S|^{2},
\label{Omega_t4}
\eeq
where the sum reads
\bea
S = \dfrac{\mathrm{c}+1}{2} \left( f_{\Phi}(-\xi,i \alpha) + f_{\Phi}(-\xi, -i \alpha) \right) -
\dfrac{\mathrm{c}-1}{2} \exp \left(-i\dfrac{\pi}{4} \right) \left( f_{\Phi}(\xi,\alpha) - f_{\Phi}(\xi, -\alpha) \right),
\label{pi_four}
\eea 
and the phase angle associated with the structure functions is given by $\Phi = \varphi + \phi_{1}, 
\phi_{1}= (\delta -1 +2 \kappa) \frac{\pi}{4}$. With the choice of the parameters $\mathrm{c} = 1, \delta = 1$ the single
photon-added case in the tomogram 
(\ref{Omega_t4}) becomes identical to its initial state (\ref{Omega_t0}) rendering the corresponding time period assume the value $T = \frac{\pi}{4 \lambda}$. This may be observed in Fig. \ref{fig_OT_mpa} $(\mathsf{a}_{4})$.

\par

Explicit evaluation of the tomogram $\Omega(X,\varphi;t)$ at other submultiples of the time period proceeds similarly. As an 
example, we demonstrate this for the scaled time  $\lambda t=\frac{\pi}{8}$. For this purpose we employ the following identity:
\bea
\exp\left( -i \dfrac{\pi}{8} n^{2} \right) &=& \dfrac{1+(-1)^n}{2} \left\lgroup \exp\left( -i \dfrac{\pi}{2} \right) \dfrac{1-i^n}{2} 
+ \dfrac{1+i^n}{2} \right\rgroup \nn \\ 
&+&   \dfrac{1-(-1)^n}{2} \exp\left( -i \dfrac{\pi}{8} \right) 
\left\lgroup \exp\left( -i \dfrac{\pi}{4} n \right) \dfrac{1-i^{n+1}}{2} + 
\exp\left( i \dfrac{\pi}{4} n \right) \dfrac{1+i^{n+1}}{2}  \right\rgroup
\label{n_linear_pi_eight}
\eea
that allows us to establish the corresponding tomogram in the form given below:
\beq
\Omega\Big(X, \varphi; \lambda t=\frac{\pi}{8}\Big) = 
(\mathcal{N}^{(\kappa)})^{2}\; \dfrac{\exp(-X^{2})}{\sqrt{\pi}\mu 2^{\kappa}} \;
\exp \left(-|\alpha|^{2}-\frac{\alpha^{2}\nu^{*}}{2\mu}-\frac{\alpha^{*2}\nu}{2\mu} \right) |\widetilde{S}|^{2}.
\label{To8}
\eeq
The closed form expression of the sum over an infinite number of modes (\ref{OTt}) for the chosen time 
$\lambda t=\frac{\pi}{8}$ may be furnished as a linear combination of structure functions:
\bea
\widetilde{S} \!\!\! &=& \!\!\! \dfrac{\mathrm{c}+1}{4} \left \lgroup (1-i) \left( f_{\widetilde{\Phi}}(\xi,\alpha)
+ f_{\widetilde{\Phi}}(\xi, -\alpha) \right)  + (1+i) \left( f_{\widetilde{\Phi}}(-\xi,i \alpha)
+ f_{\widetilde{\Phi}}(-\xi, -i\alpha) \right) \right \rgroup  + \dfrac{\mathrm{c}-1}{4} \exp \left(i \dfrac{\pi}{8} \right) 
\times \qquad \nn \\
&&\times  \left \lgroup (1-i) \left( f_{\widetilde{\Phi}+\frac{\pi}{4}}(\xi,- \alpha) 
- f_{\widetilde{\Phi}+\frac{\pi}{4}}(\xi, \alpha) \right)  + 
(1-i) \left( f_{\widetilde{\Phi}-\frac{\pi}{4}}(\xi,- \alpha) 
- f_{\widetilde{\Phi}-\frac{\pi}{4}}(\xi, \alpha) \right)  \right \rgroup, \quad
\label{pi_eight}
\eea
where the effective phase contribution reads $ \widetilde{\Phi} = \varphi + \phi_{2}, \phi_{2}=(\delta -1 +2\kappa) \frac{\pi}{8}$.
As remarked earlier, the square of the modulus of each structure function in (\ref{pi_eight}) produces a strand in the 
corresponding tomogram 
($\mathrm{c}=1, \kappa = 1,\delta = 1$ case) given in Fig. \ref{fig_OT_mpa} $(\mathsf{a}_{3})$. The bilinear interference terms produced 
by the cross multiples of the structure functions give rise to the striations in the 
tomogram. These features are generic at all rational submultiples of the time period, and, in particular, may also be observed for the 
preceding set of parametric values in
Fig.  \ref{fig_OT_mpa} $(\mathsf{a}_{2})$, which  represents the tomogram  $\Omega(X, \varphi; \lambda t= \frac{\pi}{12})$.  

\subsection{Nonclassical depth}
\label{NonClaDe}
To investigate the extent of nonclassicality of an arbitrary quantum state a general phase space distribution function depending 
on a continuous real variable $\sigma$ has been introduced [\cite{L1991}, \cite{L1992}] as
\beq
R(\beta,\beta^{*};\sigma)=\dfrac{1}{\pi \sigma} \int P(\gamma,\gamma^{*}) 
\exp \Big( -\dfrac{|\beta - \gamma|^{2}}{\sigma} \Big)\, \mathrm{d}^{2}\gamma,
\label{ncdepth}
\eeq
where the distribution $R(\beta, \beta^{*};\sigma)$ embodies a smoothing process induced by a Gaussian convolution with a 
dispersion $\sigma$. The $R$-distribution  satisfies the normalizability condition for an arbitrary $\sigma$:
\beq
\int R(\beta,\beta^{*};\sigma)\, \mathrm{d}^{2}\beta = 1.
\label{R_norm}
\eeq
 For the example considered here the distribution $R(\beta, \beta^{*};\sigma)$ interpolates in the domain 
$\sigma \in [0, 1]$ between the highly singular $P$-representation (\ref{P_equation_mpa}) and the positive semidefinite 
$Q$-function (\ref{Q_function_pa}). The greatest 
lower bound of the smoothing parameter $\sigma$ that renders the $R$-distribution of a 
relevant quantum state positive semidefinite may be regarded [\cite{L1991}] as its measure of nonclassicality. Using this procedure 
the nonclassical depths of various states have been investigated [\cite{L1991}-\cite{HWLZ2016}]. Towards evaluating the 
distribution $R(\beta, \beta^{*};\sigma)$ for the quantum state studied here we substitute the $P$-representation 
(\ref{P_equation_mpa}) in the definition (\ref{ncdepth}). Utilizing the identity (\ref{Int_hyper}) we obtain
\bea
R^{(\kappa)}(\beta, \beta^{*};\sigma;t) \!\!\!\! &=& \!\!\!\! \dfrac{1}{\pi \sigma}
\sum_{n,m=\kappa}^{\infty} \!  \dfrac{\sigma^{-(n+m)}}{\sqrt{n!m!}} 
  \beta^{*n} \beta^{m}  \exp \left(-\dfrac{|\beta|^{2}}{\sigma} \right)\,{}_2F_0\left( \! \! -n,-m;\phantom{}_{-} ;
-\frac{\sigma(1-\sigma)}{|\beta|^{2}}\right) \dm(t).
\label{ncdepth_mpa}
\eea
Owing to the conservation of trace of the density matrix the construction (\ref{ncdepth_mpa}) maintains the normalizability criterion (\ref{R_norm}). Employing the explicit values of the density matrix elements (\ref{den_nm}) we notice that at times which are rational 
submultiples of the period of the Wehrl entropy, the double summation in (\ref{ncdepth_mpa}) may be reduced to a single 
infinite sum. To achieve this we use the following identity:
\bea
\sum_{n=0}^{\infty}\dfrac{(n+ \ell)! \, \mathsf{t}^{n}}{n! \, (n+ \ell-\kappa)!} \, \mathrm{H}_{n+\ell-\kappa}(\mathsf{x})
 &=& 
\exp(-\mathsf{t}^{2}+2 \, \mathsf{x} \, \mathsf{t})
\sum_{p=0}^{\kappa} \, \mathsf{t}^{p} \, (\kappa -p)! \, \binom{\ell}{\kappa -p} \binom{\kappa}{p}  
\mathrm{H}_{\ell -\kappa +p}(\mathsf{x}-\mathsf{t}).
\label{Hermite_identity_R}
\eea
The initial $t=0$ limit of the $R$-distribution may now be readily furnished as
\beq
R^{(\kappa)}(\beta, \beta^{*};\sigma;t=0) =
\dfrac{(\mathcal{N}^{(\kappa)})^{2}}{\pi \sigma}  \exp\left (-\dfrac{|\beta|^2}{\sigma} \right )
\sum_{\ell=0}^{\infty} \dfrac{1}{\ell !} \left( \dfrac{\sigma(\sigma-1)}{ |\beta|^{2}} \right)^{\ell} \,
\Big| \mathcal{G}_{\ell}(\xi,\alpha)+ \mathrm{c} \, \mathcal{G}_{\ell}(\xi,-\alpha)\Big|^{2},
\label{R_t0}
\eeq
where the mode contribution reads
\bea
 \mathcal{G}_{\ell}(\xi,\alpha) &=& \dfrac{1}{\sqrt{\mu}} \exp \left(-\frac{|\alpha|^{2}}{2}-\frac{\alpha^{2}\nu^{*}}{2\mu}
 +\dfrac{\beta^{*2}\nu}{2\sigma^2 \mu}+ \dfrac{\alpha \beta^{*}}{\sigma \mu}\right) 
 \Big(i \sqrt{\dfrac{\nu}{2\mu}}\Big)^{\ell-\kappa} \Big(\dfrac{\beta^*}{\sigma}\Big)^{\ell} \times \nn \\
 && \times \sum_{p=0}^{\kappa} \Big(\dfrac{i\beta^*}{\sigma} \sqrt{\dfrac{\nu}{2\mu}}\Big)^p (\kappa-p)!\binom{\ell}{\kappa-p}
 \binom{\kappa}{p}\mathrm{H}_{\ell-\kappa +p} \left(\dfrac{-i\alpha \sigma - i \beta^* \nu}{\sigma \sqrt{2 \mu \nu }}\right).
 \label{G_l}
\eea
A parallel derivation that employs the identity (\ref{n_linear_pi_four}) provides the $R$-distribution at
$\lambda t=\frac{\pi}{4}$:
 \beq
R^{(\kappa)}\left(\beta, \beta^{*};\sigma;\lambda t=\frac{\pi}{4}\right) =
\dfrac{(\mathcal{N}^{(\kappa)})^{2}}{\pi \sigma}  \exp\left (-\dfrac{|\beta|^2}{\sigma} \right )
\sum_{\ell=0}^{\infty} \dfrac{1}{\ell !} \left( \dfrac{\sigma(\sigma-1)}{ |\beta|^{2}} \right)^{\ell}
\lvert \mathsf{G}_{\ell}\rvert^{2},
\label{R_pi4}
\eeq
where  the mode measure $\mathsf{G}_{\ell}$ is given by 
\bea
\mathsf{G}_{\ell} &=& \dfrac{\mathrm{c}+1}{2} \lvast \lgroup \mathcal{G}_{\ell} 
\svast( \!\!\! -\xi \exp(-2i \phi_{1}),i \alpha \exp(-i \phi_{1}) \svast)
+ \mathcal{G}_{\ell} \svast( \!\!\! -\xi \exp(-2i \phi_{1} ), -i \alpha \exp(-i \phi_{1}) \svast)
 \lvast \rgroup -
\dfrac{\mathrm{c}-1}{2} \times \qquad \quad \nn \\
&& \times \exp \left(-i\dfrac{\pi}{4} \right) \lvast \lgroup \mathcal{G}_{\ell} \svast(\xi \exp(-2i \phi_{1}),\alpha \exp(-i \phi_{1}) \svast)
- \mathcal{G}_{\ell} \svast( \xi \exp(-2i \phi_{1}), -\alpha \exp(-i \phi_{1}) \svast) \lvast \rgroup.
\label{G_pi_four}
\eea 
The expressions (\ref{R_t0}, \ref{R_pi4}) indicate that for the choice $\delta = 1, \mathrm{c} =1, \kappa = 1$, the 
$R$-distribution, following the pattern observed for the Wehrl entropy and the tomogram discussed earlier,  maintains a 
period $T= \frac{\pi}{4 \lambda}$. Proceeding similarly, we may extract the 
$R$-distribution at other rational submultiples of the period. For instance, aided by the identity (\ref{n_linear_pi_eight}),
we compute the $R$-distribution at $\lambda t=\frac{\pi}{8}$:  
\beq
R^{(\kappa)} \left(\beta ;\sigma;\lambda t=\frac{\pi}{8} \right) =
\dfrac{(\mathcal{N}^{(\kappa)})^{2}}{\pi \sigma}  \exp\left (-\dfrac{|\beta|^2}{\sigma} \right )
\sum_{\ell=0}^{\infty} \dfrac{1}{\ell !} \left( \dfrac{\sigma(\sigma-1)}{ |\beta|^{2}} \right)^{\ell} \,
\lvert \widetilde{\mathsf{G}}_{\ell} \rvert^{2},
\label{R_pi8}
\eeq
where the entry $\widetilde{\mathsf{G}}_{\ell}$  is listed as follows
\bea
\widetilde{\mathsf{G}}_{\ell} &=& \dfrac{\mathrm{c}+1}{4} \lvast \lgroup (1-i) \svast \lgroup \mathcal{G}_{\ell}
\svast(\xi\exp(-2 i \phi_{2}),\alpha\exp(-i \phi_{2}) \svast)
+ \mathcal{G}_{\ell} \svast(\xi\exp(-2i \phi_{2}), -\alpha \exp(-i \phi_{2}) \svast) \svast \rgroup  \nn \\
&+ &  \!\!\!\!\!  (1+i) \svast \lgroup \mathcal{G}_{\ell} \svast( \!\!\! -\xi\exp(-2i \phi_{2}),i \alpha \exp(-i \phi_{2})\svast)
+ \mathcal{G}_{\ell} \svast(\!\!\!-\xi \exp(-2 i \phi_{2}), \!\!\! -i\alpha \exp(-i \phi_{2})\svast) \svast \rgroup \lvast \rgroup
 \nn \\
& - & \!\!\!\!\!  \dfrac{\mathrm{c}-1}{4} \exp \left(i \dfrac{\pi}{8} \right) 
(1-i)  \lvast \lgroup   \mathcal{G}_{\ell} \svast (\xi \exp \left (-2 i \left(\phi_{2}+\frac{\pi}{4} \right) \right) ,
 \alpha \exp \left (- i \left(\phi_{2}+\frac{\pi}{4} \right) \right) \svast) \nn \\
&-&  \mathcal{G}_{\ell} \svast (\xi \exp \left (-2 i \left(\phi_{2}+\frac{\pi}{4} \right) \right) ,
- \alpha \exp \left (- i \left(\phi_{2}+\frac{\pi}{4} \right) \right) \svast) \nn \\
&+& \!\!\!\!\! 
  \mathcal{G}_{\ell} \svast (\xi \exp \left (-2 i \left(\phi_{2}-\frac{\pi}{4} \right) \right) ,
 \alpha \exp \left (- i \left(\phi_{2}-\frac{\pi}{4} \right) \right) \svast) \nn \\
&-&  \mathcal{G}_{\ell} \svast (\xi \exp \left (-2 i \left(\phi_{2}-\frac{\pi}{4} \right) \right) ,
- \alpha \exp \left (- i \left(\phi_{2}-\frac{\pi}{4} \right) \right) \svast)  \lvast \rgroup.
\label{R_pi_eight}
\eea
\begin{figure}[H]
	\captionsetup[subfigure]{labelformat=empty}
	\subfloat[]{\includegraphics[width=7cm,height=5cm]
		{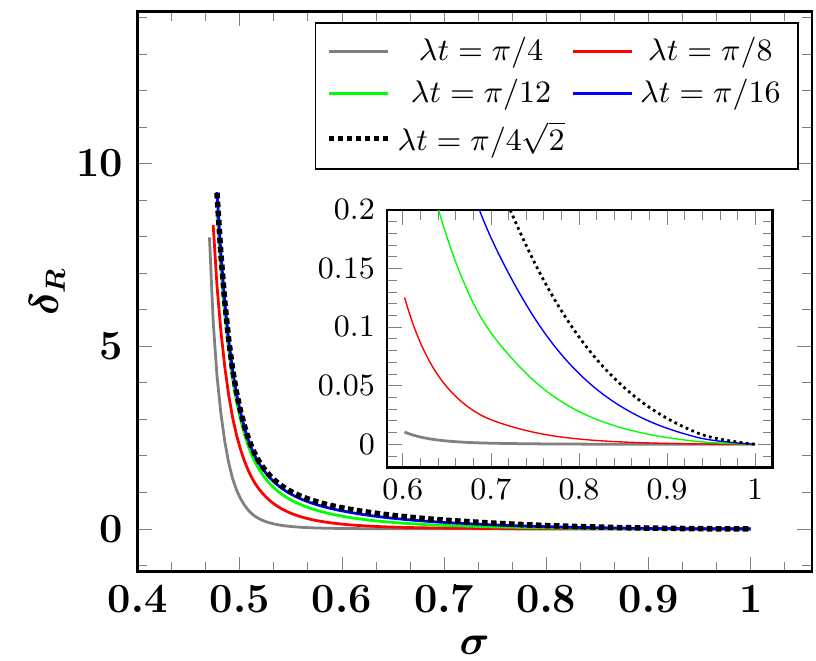}}
	\caption{The negativity $\delta_{R}$ of the $R$-distribution is plotted for the evolution of the initial state 
	(\ref{initial-state-mpa}) at various times for the choice $\mathrm{c}=1, \kappa=1, \delta = 1$, while maintaining the displacement and squeezing parameters  $\alpha=2, \xi=0.5$, respectively.}
	\label{fig_nc_depth}
\end{figure}
\par

As the expressions of the $R$-distribution obtained above contain one or more infinite summations, it becomes difficult to procure 
an analytical solution of the nonnegativity constraint $R^{(\kappa)} \big(\beta, \beta^{*};\sigma;t\big) \ge 0$ satisfied by the 
least value of $\sigma$. To proceed, we mimic the well-known description of negativity of the $W$-distribution advanced in 
[\cite{KZ2004}]. One way of numerically estimating  the extent of nonclassicality of the quantum state 
(\ref{density_matrix_mpa}) is to introduce a measure of the negative volume of the $R$-distribution for the whole range $[0, 1]$ 
of  the variable smoothing parameter $\sigma$:
\beq 
\delta_{R}(\sigma) = \int |R(\beta, \beta^{*};\sigma)|\, \mathrm{d^2}\beta-1.
\label{negativity_nc}
\eeq
The least value of the parameter $\sigma_{L}$ yielding a nonnegative phase space volume of the $R$-distribution  
$\delta_{R}(\sigma \ge \sigma_{L}) = 0$ may now be 
regarded as a measure of nonclassicality of the state. In Fig. \ref{fig_nc_depth} we observe that 
$\delta_{R}(\sigma) \rightarrow 0$ 
is realized only as $\sigma \rightarrow 1$ \textit{i.e.} when the $R$-distribution coincides with the $Q$-function. A general 
feature emerging in Fig. \ref{fig_nc_depth} is that in the range $0 < \sigma <1$ the states with a greater number of 
kitten-like structures (say, at $\lambda t=\frac{\pi}{8}$ in comparison with $\lambda t=\frac{\pi}{4}$) also possess a higher degree 
of negativity $\delta_{R}(\sigma)$. The expressions (\ref{R_pi4}, \ref{R_pi8}) indicate that for the latter 
the existence of a larger number of
interference terms between various mode contributions $\mathcal{G}_{\ell}$ gives rise to more oscillatory nature of the corresponding 
distribution $R^{(\kappa)} \left(\beta ;\sigma; \lambda t=\frac{\pi}{8} \right)$, which concomitantly now engenders a higher 
negativity at $ \lambda t=\frac{\pi}{8}$.
Another characteristic becomes evident in Fig. \ref{fig_nc_depth}.  At times such as $\lambda t = \frac{\pi}{4 \sqrt{2}}$ which are 
irrational multiples of the period, the kitten-like formations are not realized 
in the phase space. These states, however, portray consistently large negativity measure $\delta_{R}(\sigma)$ as they are 
characterized by rapidly oscillatory $R$-distribution.
\section{Decoherence models}
\label{decoherence}
\setcounter{equation}{0}
In general, the open systems [\cite{CL1983}, \cite{WCM1985}] that admit interactions with suitable environmental  degrees of freedom are characterized by the irreversible loss of information and the dissipative processes. The Lindblad 
 master equation [\cite{L1976}, \cite{GFVKS1978}]  for the time evolution of the density matrix provides a simple framework for a dynamical  
 description of the decoherence mechanism. In the context of the evolution of a quantum state in a Kerr medium interacting with a 
reservoir at zero temperature the Lindblad master equation assumes [\cite{PB2002}] the form 
\beq
\dfrac{\mathrm{d}\rho}{\mathrm{d}t}= -i[H,\rho] + \gamma \, \left( [\mathcal {X}\rho, \mathcal{X}^{\dagger}]
+[\mathcal{X},\rho \mathcal{X}^{\dagger}] \right),
\label{LM_equation}
\eeq
where $\mathcal{X}$ is a suitable  operator embodying the dissipation mechanism, and $\gamma$ is the damping factor. Employing the algebraic structure of the superoperators an analytical solution to the master equation (\ref{LM_equation}) has been provided in 
[\cite{VSM2016}]. In our analysis we closely follow the method prescribed therein. 
\subsection{Amplitude decay model}
In the Born-Markov approximation at zero temperature 
the master equation (\ref{LM_equation}) relevant for the amplitude decay process for the Kerr Hamiltonian (\ref{KerrH}) is given by
the choice $\mathcal{X} = a$.
Following [\cite{VSM2016}] we introduce the superoperators acting on an arbitrary density matrix $\rho$ associated with the 
system interacting with the reservoir degrees of freedom.  In the Heisenberg picture the action of the superoperators 
may be enlisted as
\beq
\mathsf{S} \rho \equiv -i [H, \rho],\;\; \mathsf{J} \rho \equiv 2 \gamma a \rho a^{\dagger},\;\;
\mathsf{L} \rho \equiv - \gamma \big(a^{\dagger} a \rho + \rho a^{\dagger} a\big),\;\;
\mathsf{R} \rho \equiv a^{\dagger} a \rho - \rho a^{\dagger} a. 
\label{suop}
\eeq
The algebraic structure satisfied by the superoperators reads
\beq
[\mathsf{S}, \mathsf{J}] = 2 i \lambda \mathsf{R}\, \mathsf{J},\;\; [\mathsf{S}, \mathsf{L}] = 0, \;\;
[\mathsf{S}, \mathsf{R}] = 0,\;\; [\mathsf{L}, \mathsf{J}] = 2  \gamma \mathsf{J},\;\; [\mathsf{L}, \mathsf{R}] = 0, \;\;
[\mathsf{R}, \mathsf{J}] = 0.
\label{sucom}
\eeq
The above commutation relations admit [\cite{VSM2016}] a factorized form of the evolution of the density matrix of the dissipative 
system. In particular, the amplitude decay process now restricts the initial state (\ref{initial-state-mpa}) to evolve
to the mixed state density matrix given below: 
\beq
\rho^{(\kappa)}(t)  = \exp(\mathsf{S}\, t)\; \exp(\mathsf{L}\, t)\; \exp\left\lgroup 
\frac{1 - \exp\big(-2(\gamma + i \lambda \mathsf{R}) t\big)}{2(\gamma + i \lambda \mathsf{R})}\, \mathsf{J}\right\rgroup\;
\rho^{(\kappa)}(0).
\label{den-Lin-t}
\eeq
The above evolution equation in conjunction with the operator structure (\ref{suop}) now readily furnishes  the amplitude damped 
elements of the density matrix (\ref{density_matrix_mpa}) in the number state basis: 
\bea
\dm(t) &=& \exp\big(-i\omega(n-m)t\big)\, \exp\big(-i\lambda(n-m)(n+m-1)t\big) \, \exp\big(-\gamma(n+m)t\big) \, \times \nn \\
&& \times \sum_{\ell=0}^{\infty} \left( \dfrac{(n+\ell)!(m+\ell)!}{n!m!}\right)^{\tfrac{1}{2}} 
\dfrac{1}{\ell!} \left( \dfrac{\gamma (1-\exp(-2(\gamma+i\lambda(n-m)) t ))}{(\gamma+i\lambda(n-m))}\right)^{\ell} \numdm(0).
\label{density_amp_decay}
\eea
The boundary values of the matrix elements $\numdm(0)$ may be easily read off from the construction of the initial state 
(\ref{initial-state-mpa}). In (\ref{density_amp_decay}) the higher mode components of the density matrix decay more rapidly, and, 
consequently, in the long time limit only the ground state remains populated: 
$\dm(t \rightarrow \infty) \longrightarrow \delta_{n,0}\,\delta_{m,0}$.
The construction (\ref{density_amp_decay}) of the density matrix elements immediately furnishes the Husimi $Q$-function (\ref{Q_defn}) for the system undergoing the amplitude decay:
\beq
Q(\beta,\beta^{*},t)= \dfrac{1}{\pi} \exp(-|\beta|^{2}) \sum_{n,m=0}^{\infty} \dfrac{\beta^{*n}\,\beta^{m}}{\sqrt{n!\,m!}}\,
\dm(t)\xrightarrow[t \rightarrow \infty]{} \dfrac{1}{\pi} \exp(-|\beta|^{2}).
\label{Q-rho}
\eeq
Consequently, the asymptotic value of the corresponding Wehrl entropy (\ref{WehrlDef}) in the 
presence of amplitude dissipation terms in the Lindblad equation reads  $S_{Q}(t\rightarrow \infty) 
\longrightarrow 1 + \log \pi$, which is the universal lower bound of $S_{Q}$ [\cite{L1978}]. This is observed 
in Figs. \ref{fig_amp_decay} ($\mathsf{a}_{1}, \mathsf{a}_{2}, \mathsf{a}_{8}$). The decoherence phenomenon 
and the loss of nonclassicality in the presence of amplitude dissipation (\ref{density_amp_decay})  are evident from the 
asymptotic behavior of the negativity $\delta_{W}$ of the $W$-distribution (\ref{W_function_mpa}): 
$\delta_{W}(t \rightarrow \infty)  \longrightarrow 0$ (Figs. \ref{fig_amp_decay}, $\mathsf{a}_{3}, \mathsf{a}_{4}, \mathsf{a}_{9}$).
  For a more squeezed state with a larger parameter $r$ the 
higher harmonics in the density matrix (\ref{density_matrix_mpa}) are activated causing an increment in the corresponding 
$\delta_{W}$. This effect becomes more pronounced as higher harmonics induce more rapid oscillations on phase space. An increase in
$\delta_{W}$ for an enhanced squeezing parameter $r$ is observed at times $t \ll \gamma^{-1}$. However, since the higher harmonics experience increased damping, the negativity $\delta_{W}$ for a larger $r$ also 
dissipates faster (Fig. \ref{fig_amp_decay} $\mathsf{a}_{3}$) at $t \gtrsim \gamma^{-1}$. A parallel reason also induces similar behavior 
(Fig. \ref{fig_amp_decay} $\mathsf{a}_{9}$) for an increased number of photon added ($\kappa$) state. We also enlist the optical tomogram 
(\ref{optical_tomogram}) following from the evolution (\ref{density_amp_decay}) of the density matrix subject to amplitude damping:
\beq
\Omega(X,\varphi;t)=\dfrac{\exp(-X^{2})}{\sqrt{\pi}}
\sum_{n,m=0}^{\infty} \dfrac{\exp(-i\varphi(n-m))}{\sqrt{2^{n+m}n!m!}} \,
\mathrm{H}_{n}(X)\, \mathrm{H}_{m}(X)\, \dm(t).
\label{tom-amp-disp}
\eeq
The Figs. \ref{fig_amp_decay} ($\mathsf{a}_{5}-\mathsf{a}_{7}$) illustrating the tomogram (\ref{tom-amp-disp}) at various times establish 
that the higher harmonics are progressively erased due to dissipation, 
and in the $t\rightarrow \infty$ limit the tomogram (Fig. \ref{fig_amp_decay} $\mathsf{a}_{7}$) reduces to the trivial example of the 
occupation of the ground state alone: $\Omega(X,\varphi;t\rightarrow \infty)\longrightarrow \frac{1}{\sqrt{\pi}}\,\exp(-X^{2})$.
\begin{figure}[H]
	\captionsetup[subfigure]{labelformat=empty}
	\subfloat[(a$_{1}$)]{\includegraphics[width=4cm,height=3.3cm]
		{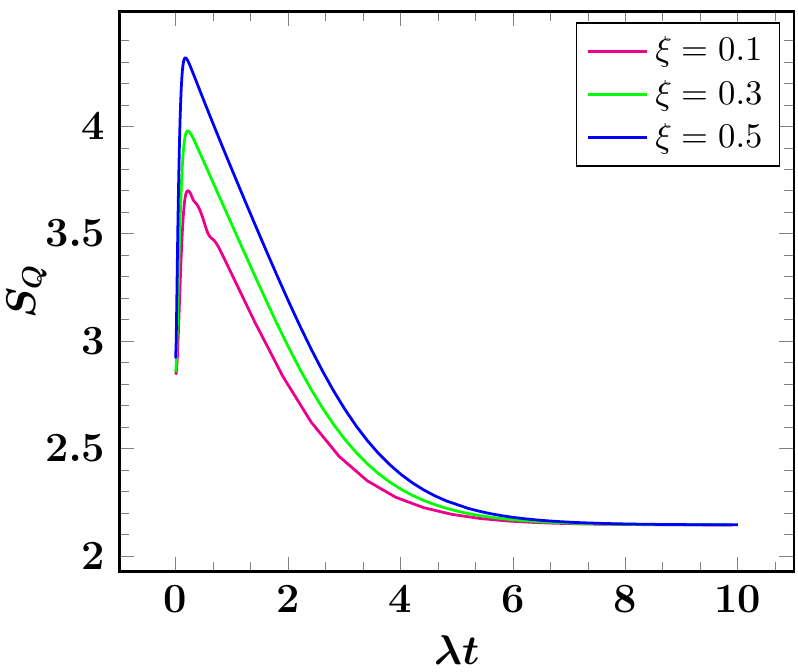}}
	\captionsetup[subfigure]{labelformat=empty}
	\subfloat[(a$_{2}$)]{\includegraphics[width=4cm,height=3.3cm]
		{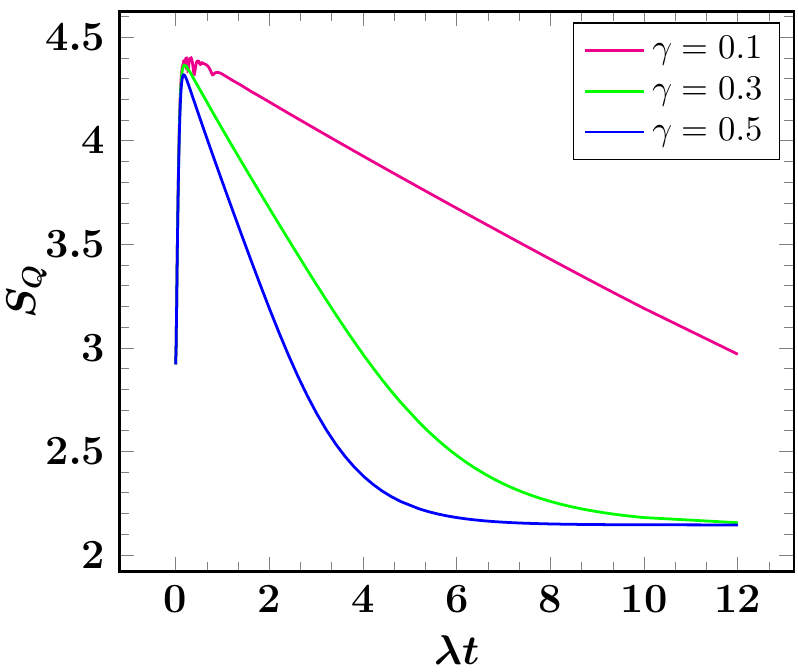}}
	\captionsetup[subfigure]{labelformat=empty}
	\subfloat[(a$_{3}$)]{\includegraphics[width=4cm,height=3.3cm]
		{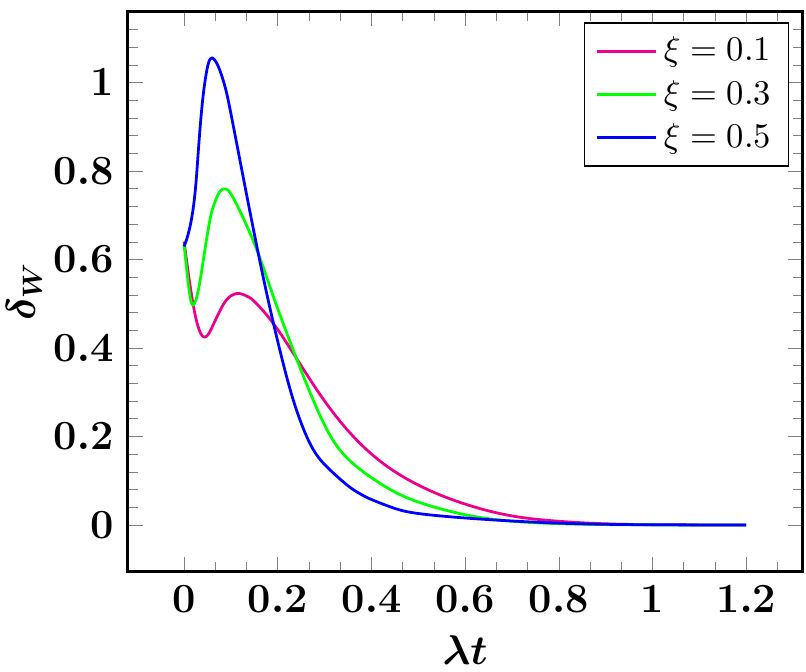}} 
	\captionsetup[subfigure]{labelformat=empty}
	\subfloat[(a$_{4}$)]{\includegraphics[width=4cm,height=3.3cm]
		{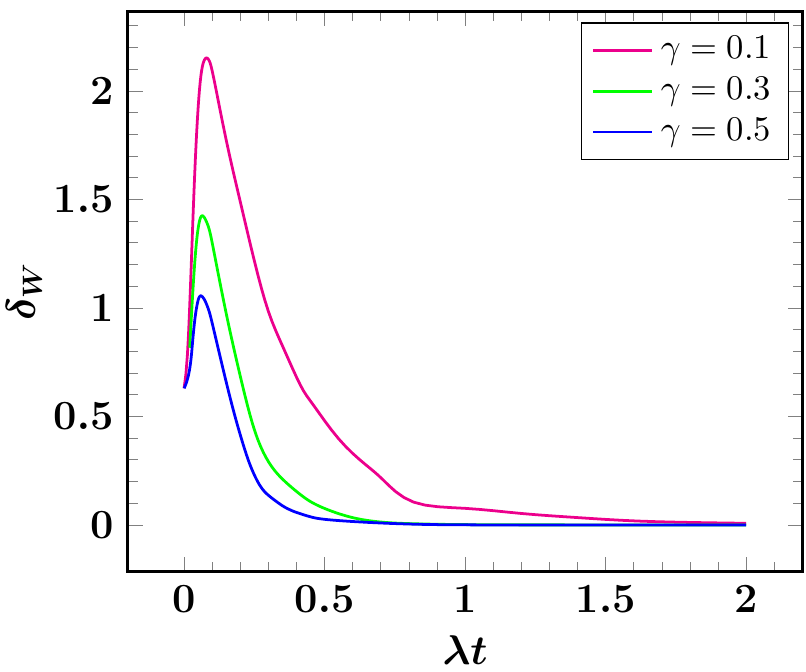}} \\
%
	\subfloat[(a$_{5}$)]{\includegraphics[scale=0.33]
		{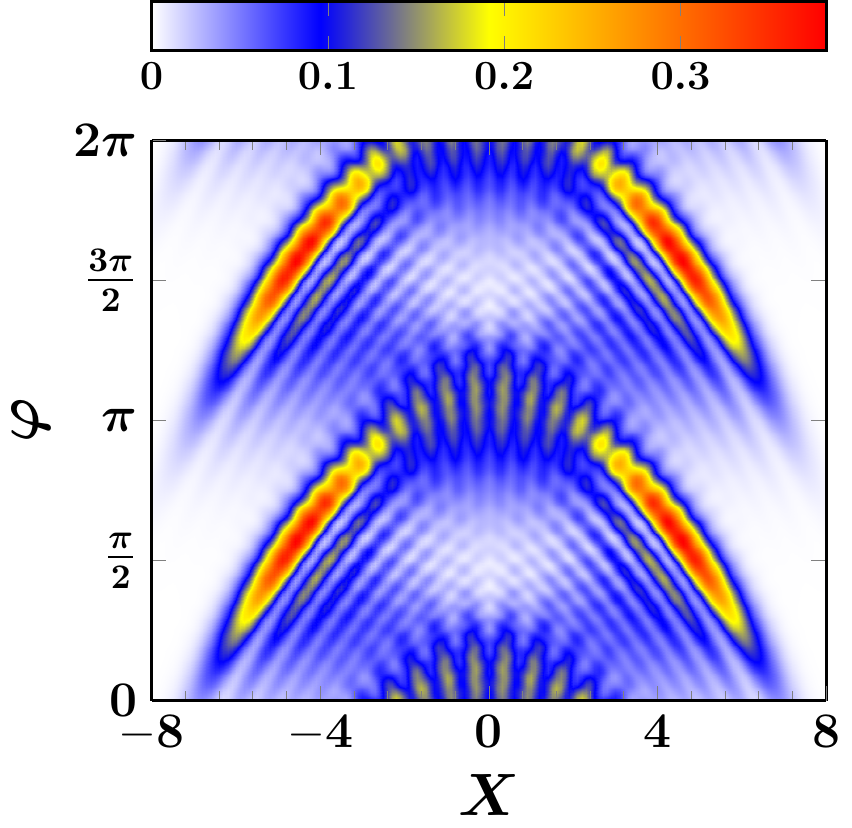}}
	\subfloat[(a$_{6}$)]{\includegraphics[scale=0.33]
		{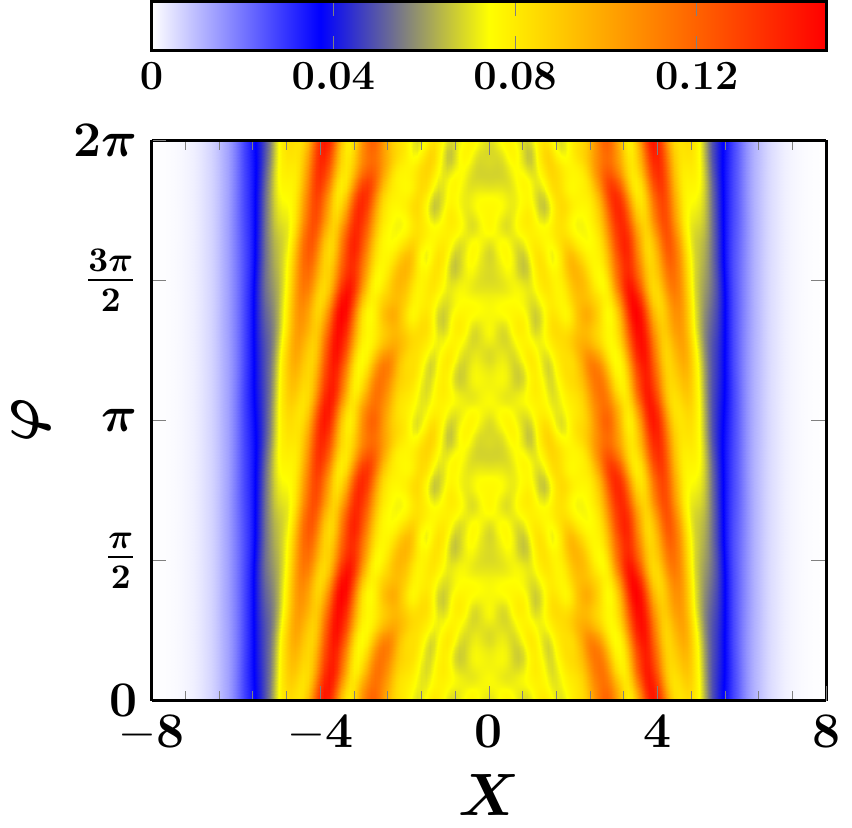}}
	\subfloat[(a$_{7}$)]{\includegraphics[scale=0.33]
		{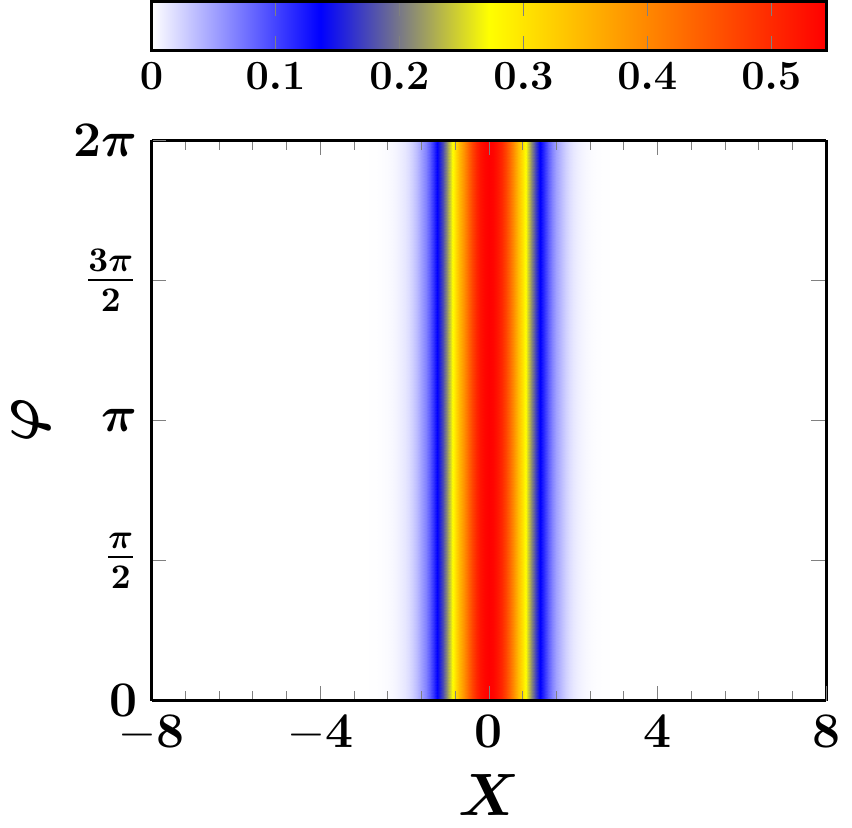}} \quad
	\subfloat[(a$_{8}$)]{\includegraphics[width=3.5cm,height=2.7cm]
		{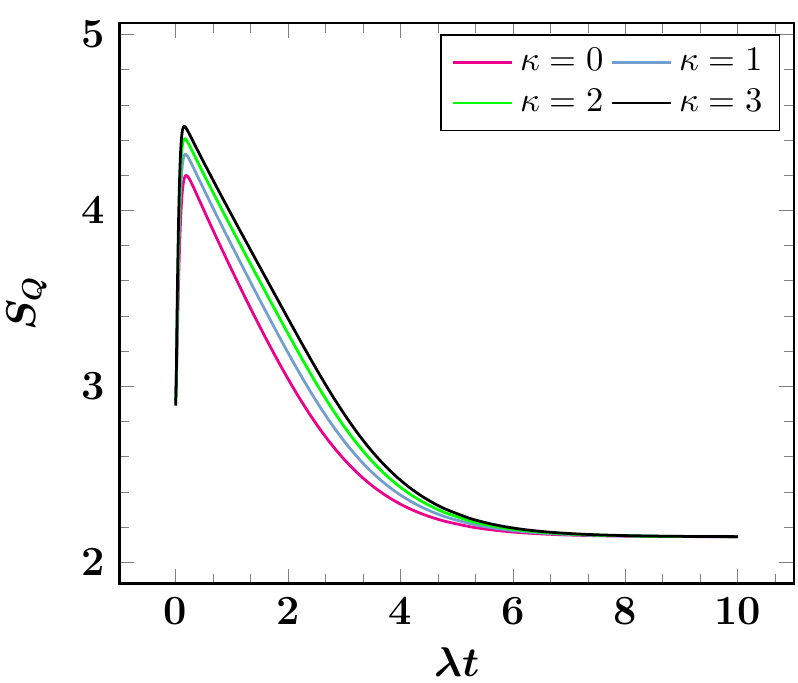}} \quad
	\subfloat[(a$_{9}$)]{\includegraphics[width=3.5cm,height=2.7cm]
		{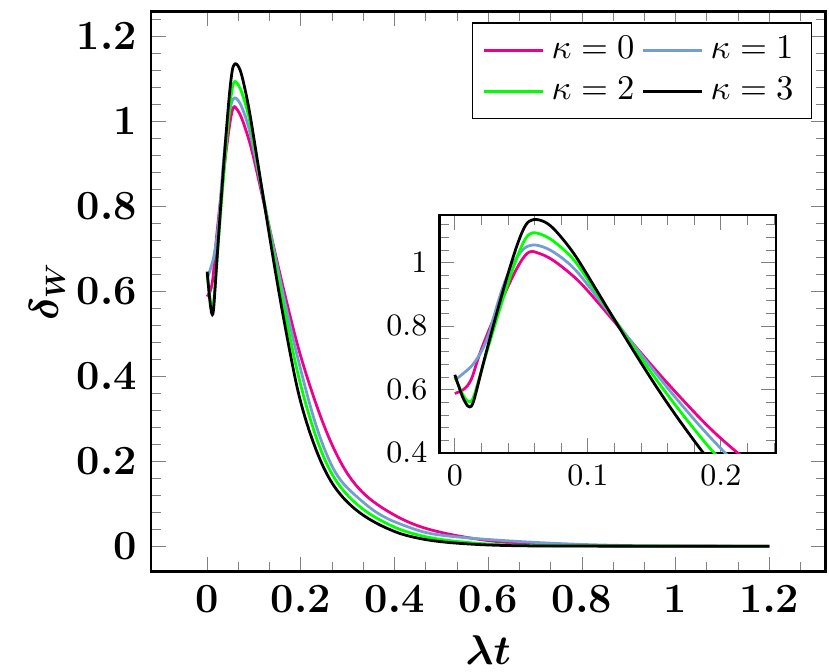}}
     \caption{For the parametric values $\mathrm{c}=1, \delta = 1, \alpha=2$ the amplitude decay model is studied. 
		Single photon-added case ($\kappa=1$) is investigated in plots ($\mathsf{a}_{1}-\mathsf{a}_{7}$). 
		The illustrations ($\mathsf{a}_{1}, \mathsf{a}_{2}$) study the evolution of the Wehrl entropy; first, 
		 for various values of the squeezing parameter $\xi$ at damping coefficient $\gamma=0.5$ ($\mathsf{a}_{1}$), and 
		 subsequently for a fixed squeezing $\xi=0.5$ and varying damping strengths ($\mathsf{a}_{2}$). 
		Similarly the depictions ($\mathsf{a}_{3}$) and ($\mathsf{a}_{4}$) refer to the evolution of negativity $\delta_{W}$ under
		the identical set of parameters  described in ($\mathsf{a}_{1}$) and ($\mathsf{a}_{2}$), respectively. 
		The diagrams ($\mathsf{a}_{5}-\mathsf{a}_{7}$) sketch the optical tomogram for  the variables $\xi=0.5$ and $\gamma=0.5$ 
		at times $\lambda t=0.06, \lambda t=0.3, \lambda t=6$, respectively. 
		Lastly, the graphs ($\mathsf{a}_{8}, \mathsf{a}_{9}$) produce  the Wehrl entropy and negativity $\delta_{W}$
		for the multiple photon-added cases ($\kappa =0-3$), where the squeezing parameter and the damping constant
		read $\xi=0.5, \gamma=0.5$.}
	\label{fig_amp_decay}
\end{figure}
\subsection{Phase damping model}
For the phase damping model the interaction between the system and the reservoir degrees of freedom is represented by the 
choice  $\mathcal{X}=a^{\dagger}a$ in the evolution (\ref{LM_equation}) of the mixed state density matrix. As the 
superoperator algebra is fully commutative in the present case the elements of the  density matrix  
 in the number state basis (\ref{density_matrix_mpa}) assume the following localized form in their Fourier indices:
\bea
\rho_{n,m}^{(\kappa)}(t) = \exp\big(-i\omega(n-m)t\big) \, \exp\big(-i\lambda(n-m)(n+m-1)t\big) \, 
\exp\big(-\gamma(n-m)^{2}t\big) \, \rho_{n,m}^{(\kappa)}(0),
\label{den_phase}
\eea
where the initial values of the elements $\rho_{n,m}^{(\kappa)}(0)$ may be procured via
(\ref{initial-state-mpa}). It is evident from the phase damping process (\ref{den_phase}) that while the magnitudes of the 
diagonal components $\rho_{n,n}^{(\kappa)}(t)$
 remain invariant, the off diagonal elements undergo attenuation with time. Moreover, the damping rapidly increases as we 
move away from the diagonal elements. In the long time $t \gg \gamma^{-1}$ limit the damped density matrix elements 
assume a diagonal form, which via (\ref{den_nm}) reads:
\beq
\dm(t \rightarrow \infty) \longrightarrow \delta_{n,m}\,\rho_{n,n}^{(\kappa)}(0) =  \delta_{n,m}\,
\left( \mathcal{N}^{(\kappa)} \right)^{2} 
\Big | (1+(-1)^{n-\kappa} \mathrm{c}) \mathcal{A}_{n,\kappa}( \xi,\alpha)\Big |^{2}.
\label{DenMaPhase}
\eeq
The  diagonal asymptotically steady state density matrix (\ref{DenMaPhase}) now immediately provides the $t \gg \gamma^{-1}$ 
limit of the quasiprobability function (\ref{Q_defn}):
\bea
Q^{(\kappa)}(\beta,\beta^{*};t \rightarrow \infty) \!\!\!\! &=& \!\!\!\! 
\dfrac{\left( \mathcal{N}^{(\kappa)} \right)^{2}}{\pi \mu} 
\exp \left(-|\alpha |^{2}-|\beta |^{2}-\frac{\alpha^{2}\nu^{*}}{2\mu}-\frac{\alpha^{*2}\nu}{2\mu} \right)
|\beta |^{2 \kappa} \times \nn \\
& & \!\!\!\! \times  
\sum_{n=0}^{\infty} (1+|\mathrm{c}|^{2}+2(-1)^{n} \mathrm{Re}(\mathrm{c})) \dfrac{1}{(n!)^{2}} 
\left( \dfrac{|\nu \beta^{2}|}{2 \mu}\right)^{n} 
\left|\mathrm{H}_{n} \left(- \dfrac{i \alpha}{\sqrt{2 \mu \nu}}\right)\right|^{2}.
\label{Q_phase_decay}
\eea
The sum involving the Hermite polynomials in (\ref{Q_phase_decay}) may be recast [\cite{C1961}] in terms of the modified 
Bessel functions in the following form that is particularly amenable for asymptotic evaluations: 
\bea
\sum_{n=0}^{\infty}\dfrac{ \mathsf{t}^{n}}{(n!)^{2}} \, \mathrm{H}_{n}(\mathsf{x}) \,\mathrm{H}_{n}(\mathsf{y}) &=& 
\mathrm{I}_{0}(2 \, \mathsf{t}) \, \mathrm{I}_{0}(4 \, \sqrt{\mathsf{x} \, \mathsf{y} \, \mathsf{t}}) +
\sum_{\ell =1}^{\infty} (-1)^{\ell} \left( \dfrac{\mathsf{x}^{2 \ell}+ \mathsf{y}^{2\ell}}{\mathsf{x}^{\ell} 
\mathsf{y}^{\ell}} \right) 
\mathrm{I}_{\ell}(2 \, \mathrm{t}) \, \mathrm{I}_{2 \ell}(4 \, \sqrt{\mathsf{x} \, \mathsf{y} \, \mathsf{t}}),
\label{Hermite_Bessel_id}  
\eea
where $\mathrm{I}_{n}(\mathsf{x}) = \sum_{\ell =0}^{\infty} \frac{1}{\ell !\, \Gamma(n + \ell + 1)}
\left(\frac{\mathsf{x}}{2}\right)^{n + 2 \ell}$. The construction (\ref{Q_phase_decay}, \ref{Hermite_Bessel_id}) along with 
the phase space integral (\ref{WehrlDef}) allows us to compute the asymptotic limit of the Wehrl entropy $S_{Q}$. The 
Figs. \ref{fig_phase_decay} ($\mathsf{a}_{1}, \mathsf{a}_{2}, \mathsf{a}_{8}$) describe the evolution of $S_{Q}$  in a phase damped system. As the initial diagonal elements of the density matrix are preserved in the dissipation process (\ref{DenMaPhase}), the 
 asymptotic limit of $S_{Q}$ achieves a steady state value. Moreover, a larger value of the squeezing parameter $r$ 
 reflecting more extensive occupation in the phase space results in a comparatively enhanced asymptotic limit of  $S_{Q}$.
 For various choices of $r$ the corresponding long time limits of $S_{Q}$ are given in Table \ref{PhaseSqueeze}. 
 Analogously, a higher value of $\kappa$ for the multiple photon-added  states is linked with an increased  phase 
 space occupation of the $Q$-function, which, in turn,  produce a concomitant increment in asymptotic values 
 of $S_{Q}$ (Table \ref{PhasePhoton}).

\par

Contrasting the amplitude damping process, the \textit{conserved} diagonal density matrix elements (\ref{DenMaPhase}) in the phase 
damping model manifest nonclassicality even in the long time 
limit: $t \gg \gamma^{-1}$. A quantitative indicator of this phenomenon is revealed by the asymptotic limit of the Wigner 
distribution
\bea
W^{(\kappa)}(\beta,\beta^{*};t \rightarrow \infty) \!\!\!\! &=& \!\!\!\! \dfrac{2}{\pi \mu}
\left( \mathcal{N}^{(\kappa)} \right)^{2}
\exp \left(-|\alpha |^{2}-2|\beta |^{2}-\frac{\alpha^{2}\nu^{*}}{2\mu}-\frac{\alpha^{*2}\nu}{2\mu} \right)
|2 \beta |^{2 \kappa} \times \nn \\
& &  \times  
\sum_{n=0}^{\infty} (1+|\mathrm{c}|^{2}+2(-1)^{n} \mathrm{Re}(\mathrm{c})) \dfrac{1}{(n!)^{2}} 
\left( \dfrac{2|\nu \beta^{2}|}{\mu}\right)^{n} 
\left|\mathrm{H}_{n} \left(- \dfrac{i \alpha}{\sqrt{2 \mu \nu}}\right)\right|^{2} \times \nn \\
& & \times  {}_2F_0\Big( -(n+\kappa),-(n+\kappa);\phantom{}_{-} ; -\dfrac{1}{4|\beta|^{2}}\Big) 
\label{W_phase_decay}
\eea
obtained via (\ref{W_function_mpa}, \ref{DenMaPhase}). The limiting value (\ref{W_phase_decay}) of the $W$-distribution now 
facilitates the computation, \textit{\`a la} (\ref{negativity}), of the negativity $\delta_{W}$  represented in 
Figs. \ref{fig_phase_decay} ($\mathsf{a}_{3}, \mathsf{a}_{4}, \mathsf{a}_{9}$). Attenuation of the off-diagonal elements 
$\rho_{n,m}^{(\kappa)}(t)\, \forall n \neq m$ 
with time initially triggers the decline in $\delta_{W}$ in the scale $t \lesssim \gamma^{-1}$, but subsequently an asymptotically 
($t \gg \gamma^{-1}$) stable value of $\delta_{W}$ is realized due to the undamped diagonal density matrix elements.
The asymptotic value of the $W$-distribution (\ref{W_phase_decay}) reveals that even though its spread in the phase space 
increases with increasing $r$, the domains with negative values of  $W^{(\kappa)}(\beta,\beta^{*};t \rightarrow \infty)$ contracts 
with the increment of the squeezing parameter, as the latter diminishes the arguments of the Hermite polynomials. Consequently, a larger value of $r$ is found to yield a reduced asymptotic value of the negativity $\delta_{W}$ 
(Fig. \ref{fig_phase_decay} $\mathsf{a}_{3}$, Table \ref{PhaseSqueeze}). To validate above arguments, we have plotted the asymptotic distribution (\ref{W_phase_decay}) for $\gamma = 0.5$, while maintaining the squeezing parameters $r = 0.05$ 
(Fig. \ref{W-asymptotic} $\mathsf{a}_{1}$) and 
$r = 0.7$ (Fig. \ref{W-asymptotic}  $\mathsf{a}_{2}$), respectively. On the other hand, an increased photon-added ($\kappa > 1$) state induces 
a small augmentation in the sign reversals in the
  $W^{(\kappa)}(\beta,\beta^{*};t \rightarrow \infty)$ distribution due to the  presence of
the increasingly higher order Laguerre polynomial ${}_2F_0\left( -(n+\kappa),-(n+\kappa);\phantom{}_{-}; - (4|\beta|^{2})^{-1}\right)$ in the infinite sum. This causes a modest increment in the asymptotic value of $\delta_{W}$ with increasing $\kappa$ 
(Fig. \ref{fig_phase_decay} $\mathsf{a}_{9}$, Table  \ref{PhasePhoton}).

\par

Finally, we enlist the asymptotic behavior of the tomogram (\ref{tom-amp-disp}) subject to the evolution (\ref{DenMaPhase}) 
characterizing the phase damping model:
\bea
\Omega^{(\kappa)}(X,\varphi; t \rightarrow \infty) &=& 
 \dfrac{\left( \mathcal{N}^{(\kappa)} \right)^{2}}{\sqrt{\pi} \,  \mu \, 2^{\kappa}} 
 \exp \left(-|\alpha |^{2}-\frac{\alpha^{2}\nu^{*}}{2\mu}-\frac{\alpha^{*2}\nu}{2\mu} - X^{2} \right)
 \sum_{n=0}^{\infty} (1+|\mathrm{c}|^{2}+2(-1)^{n} \mathrm{Re}(\mathrm{c})) \times \nn \\
&& \times \dfrac{1}{(n!)^{2}} \left( \dfrac{|\nu | }{4 \mu}\right)^{n} 
 \left( \mathrm{H}_{n + \kappa}(X) \right)^{2} 
 \left|\mathrm{H}_{n} \left(- \dfrac{i \alpha}{\sqrt{2 \mu \nu}}\right)\right|^{2}. 
\label{TomPhaseAsymp}
\eea
The tomogram $\Omega^{(\kappa)}(X,\varphi; t)$ reproduced in Figs. \ref{fig_phase_decay} ($\mathsf{a}_{5}-\mathsf{a}_{7}$) 
at incremental times 
exhibit gradual disappearance of phase relationships triggered by the dissipation process. The limiting value 
(\ref{TomPhaseAsymp}) arrived at $t \gg \gamma^{-1}$ is independent of the phase $\varphi$ of the quadrature variable. This is 
confirmed in Fig. \ref{fig_phase_decay} ($\mathsf{a}_{7}$). 
\begin{figure}[H]
	\captionsetup[subfigure]{labelformat=empty}
	\subfloat[(a$_{1}$)]{\includegraphics[width=3.8cm,height=3cm]
		{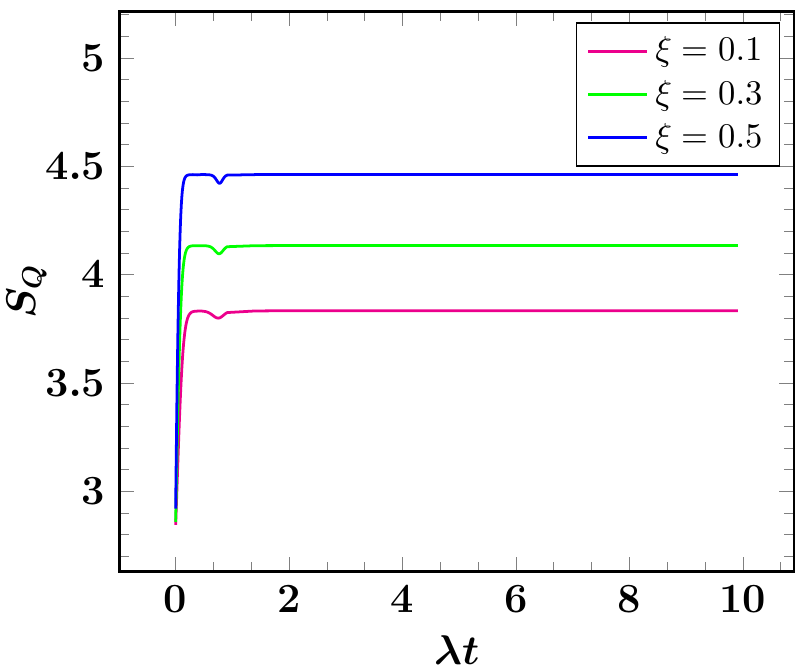}}
		\captionsetup[subfigure]{labelformat=empty}
	\subfloat[(a$_{2}$)]{\includegraphics[width=3.8cm,height=3cm]
		{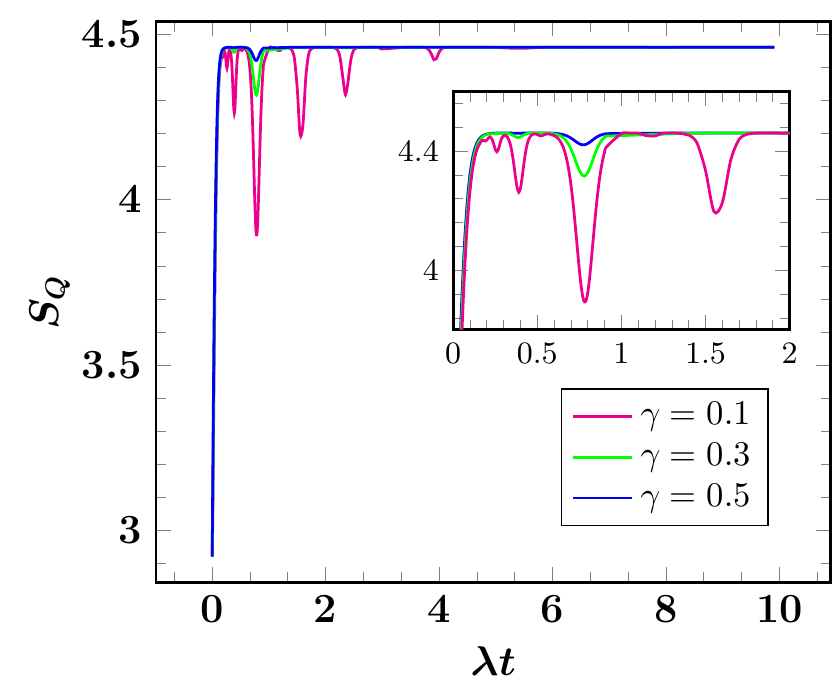}}
	\captionsetup[subfigure]{labelformat=empty}
	\subfloat[(a$_{3}$)]{\includegraphics[width=3.8cm,height=3cm]
		{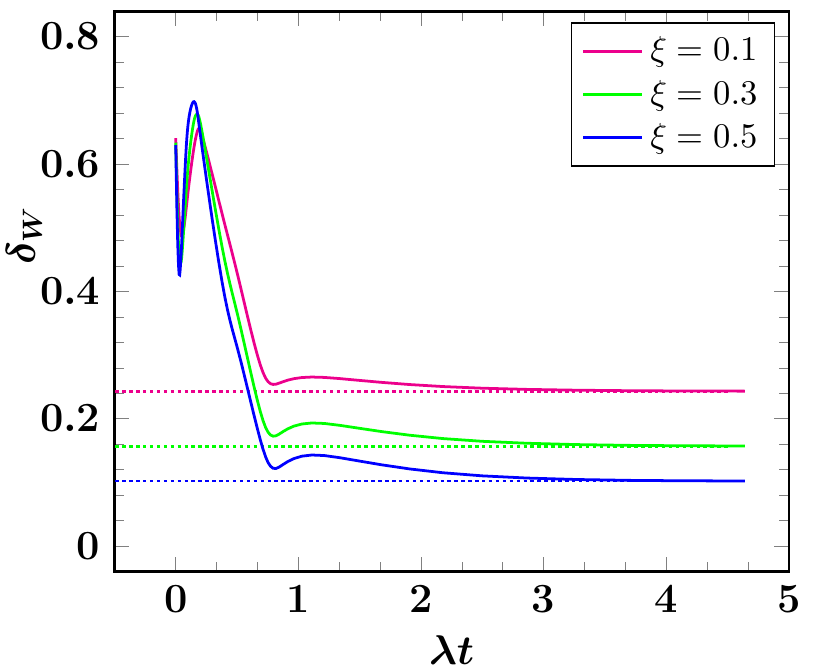}}
	\captionsetup[subfigure]{labelformat=empty}
	\subfloat[(a$_{4}$)]{\includegraphics[width=3.8cm,height=3cm]
		{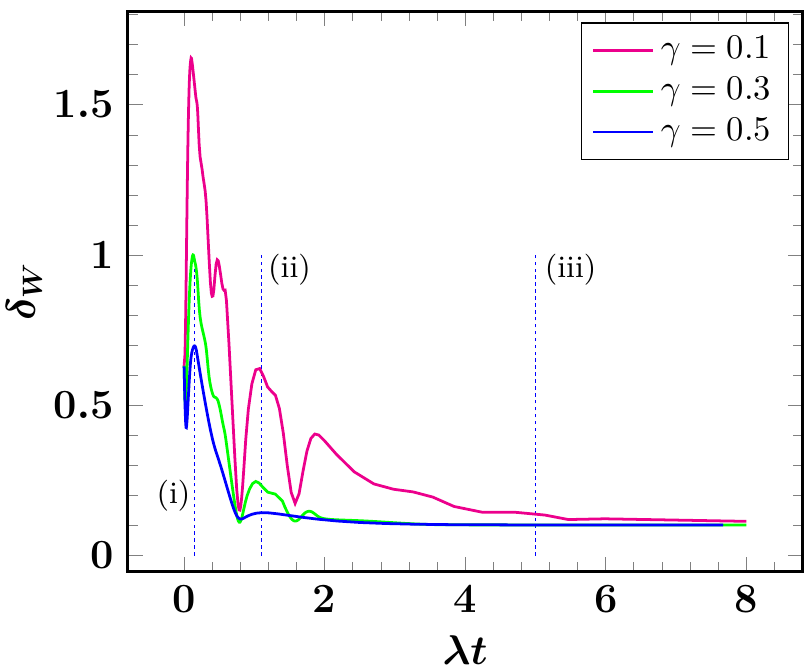}}
	\captionsetup[subfigure]{labelformat=empty}
	\subfloat[(a$_{5}$)]{\includegraphics[scale=0.35]
		{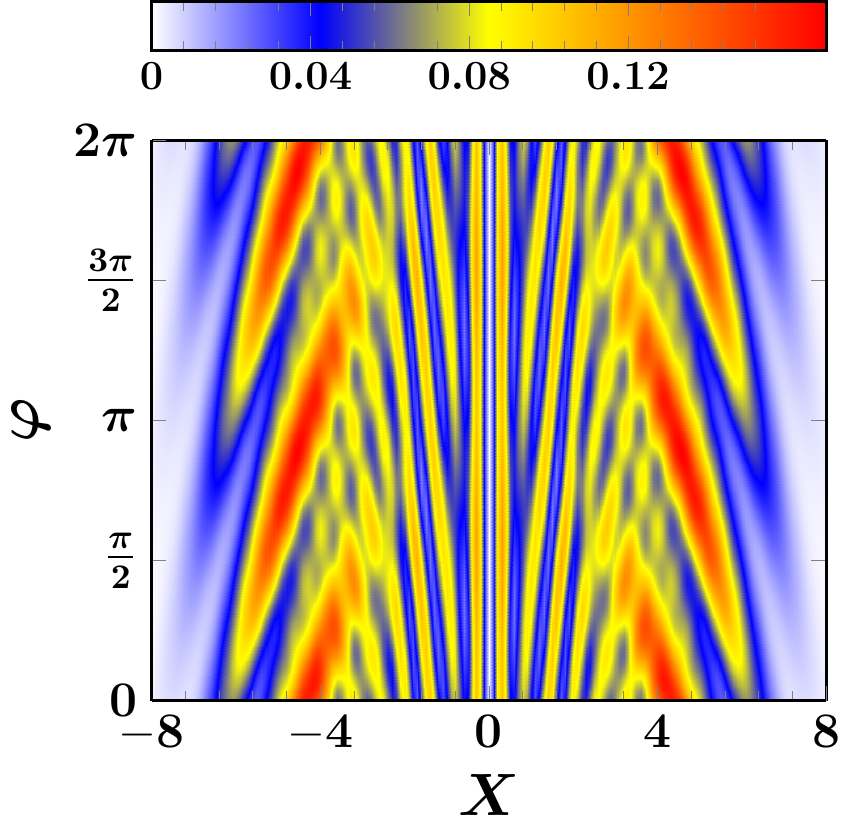}}
	\captionsetup[subfigure]{labelformat=empty}
	\subfloat[(a$_{6}$)]{\includegraphics[scale=0.35]
		{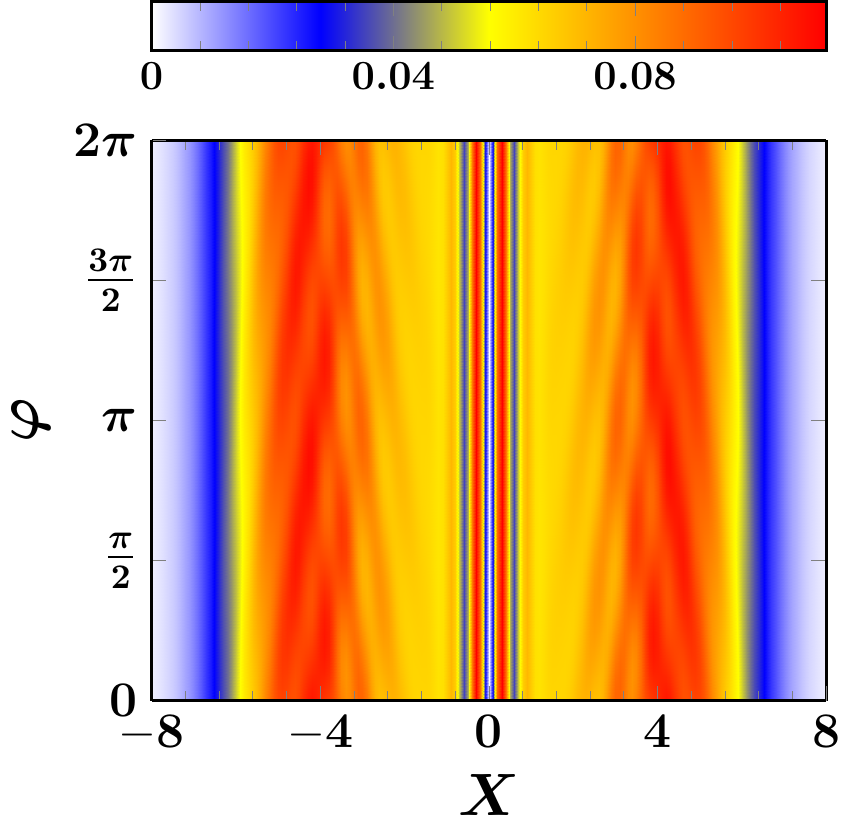}}
	\captionsetup[subfigure]{labelformat=empty}
	\subfloat[(a$_{7}$)]{\includegraphics[scale=0.35]
		{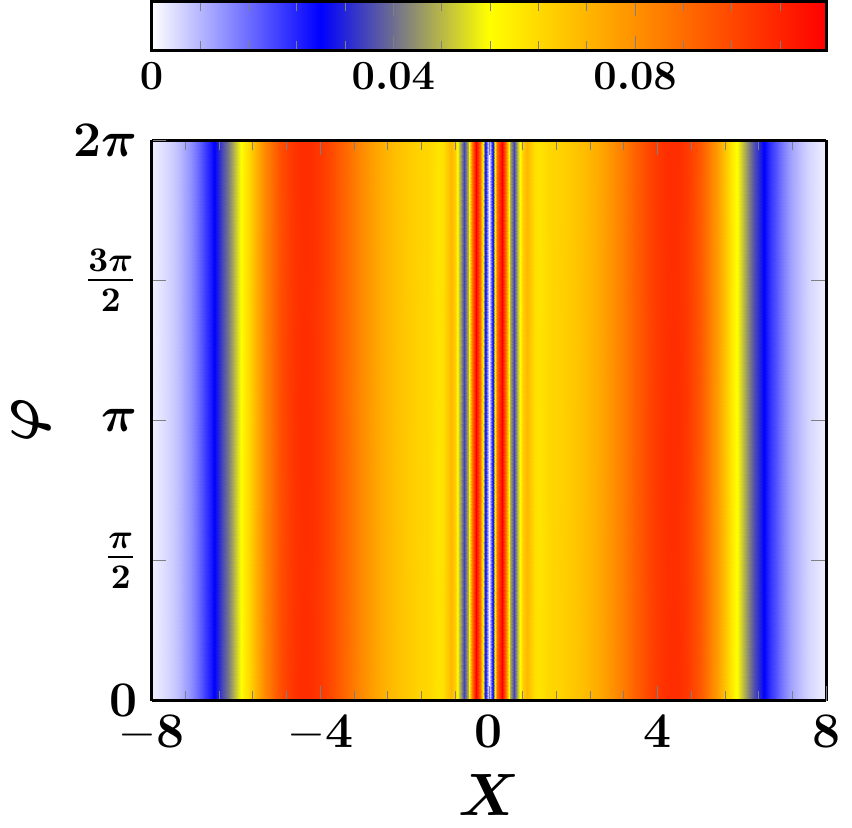}}
	\captionsetup[subfigure]{labelformat=empty}
	\subfloat[(a$_{8}$)]{\includegraphics[width=3.8cm,height=3cm]
		{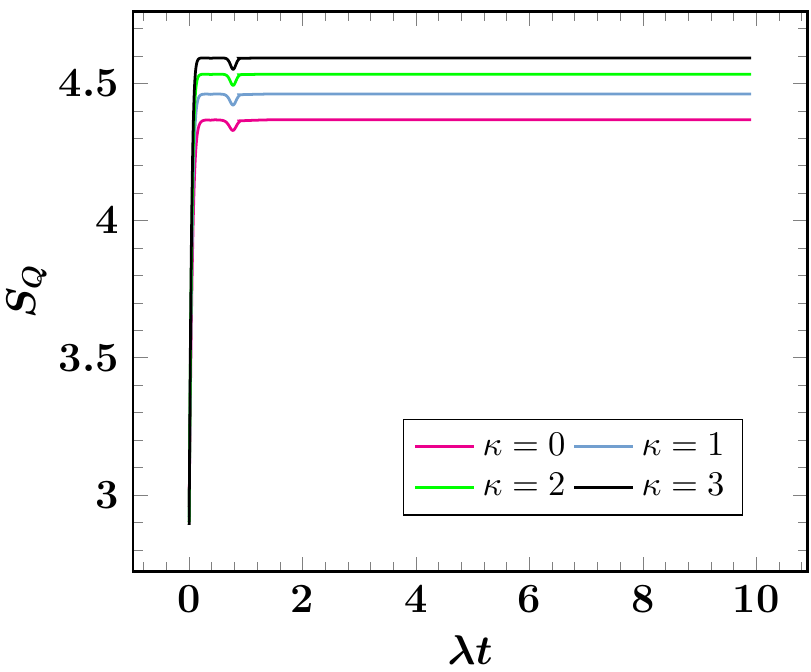}}
	\subfloat[(a$_{9}$)]{\includegraphics[width=3.8cm,height=3cm]
		{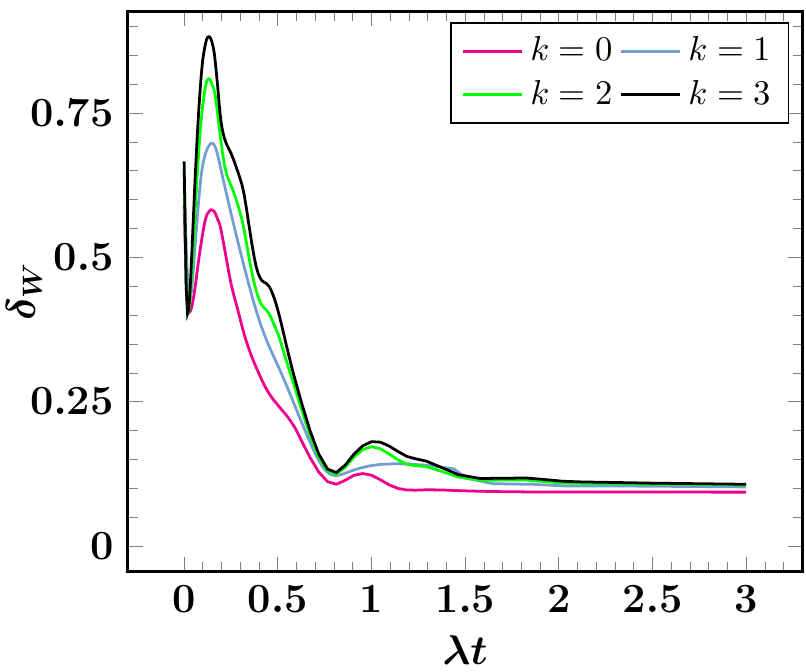}}
	\caption{Choosing the initial state with parameters $\mathrm{c}=1, \alpha=2$ the phase damping model is studied  
	for the single photon-added case $\kappa=1$ in diagrams ($\mathsf{a}_{1}-\mathsf{a}_{7}$). 
		The plots ($\mathsf{a}_{1}$) and ($\mathsf{a}_{2}$) indicate the evolution of the Wehrl entropy. In 
		($\mathsf{a}_{1}$) the  squeezing parameter $r$ is varied at damping coefficient $\gamma=0.5$, whereas ($\mathsf{a}_{2}$) probes
		varying damping strength $\gamma$ for a fixed $r=0.5$. The evolution of negativity is investigated
		in  ($\mathsf{a}_{3}$) and ($\mathsf{a}_{4}$) while maintaining  the parameters identical to those  described in 
		($\mathsf{a}_{1}$) and ($\mathsf{a}_{2}$), respectively. The asymptotic values of the $S_{Q}$ and $\delta_{W}$, observed 
		sequentially in Figs. ($\mathsf{a}_{1}$) and ($\mathsf{a}_{3}$), are given in 
		Table \ref{PhaseSqueeze} for various choices of the squeezing parameter $r$. 
		The optical tomogram of the phase damped system is explored in ($\mathsf{a}_{5}-\mathsf{a}_{7}$) at increasing times 
		$(\mathsf {i})\lambda t=0.15, (\mathsf {ii}) \lambda t=1.1, 
		(\mathsf {iii})\lambda t=5$ while the parameters $r=0.5, \gamma=0.5$ remain constants. Diagrams ($\mathsf{a}_{8},  
		\mathsf{a}_{9}$) examine the variations in the  Wehrl entropy and negativity 
		for the multiple photon-added cases ($\kappa > 1$) for  fixed values of the  parameters $r=0.5,\gamma=0.5$.
		The asymptotic values of the variables $S_{Q}$ and $\delta_{W}$ obtained in plots $\mathsf{a}_{8}$ and $\mathsf{a}_{9}$, respectively,
		are registered in Table \ref{PhasePhoton}.}
	\label{fig_phase_decay}
\end{figure}
\begin{table}[H]
\begin{center}
\begin{tabular}
{|>{\centering\arraybackslash}p{3cm}|>{\centering\arraybackslash}p{2cm} |>{\centering\arraybackslash}p{2cm}|>{\centering\arraybackslash}p{2cm}|>{\centering\arraybackslash}p{2cm}|}																								\hline
 $\kappa=1$, $\gamma=0.5$ & $r = 0.1$ & $r = 0.3$ & $r = 0.5$ \\ \cline{1-4}
 {$S_{Q} (t \rightarrow \infty)$} & $3.832996$  & $4.133899$ & $4.461696$ \\ \hline
 $  \delta_{W} (t \rightarrow \infty) $ & $0.243214$  & $0.156939$ & $0.102281$\\ \hline
\end{tabular}
\end{center}
\caption{Asymptotic ($t \gg \gamma^{-1}$) values of the Wehrl entropy ($S_{Q}$) and the negativity ($\delta_{W}$) for increasing squeezing parameter $r$ are listed following Figs. \ref{fig_phase_decay} ($\mathsf{a}_{1}$) and  ($\mathsf{a}_{3}$), respectively. The parameters and the phase space variable are fixed at $\mathrm{c} = 1, \kappa =1,\delta = 1,  \gamma = 0.5, \alpha = 2.$}
\label{PhaseSqueeze}
\end{table}
%
\begin{table}[H]
\begin{center}
\begin{tabular}
{|>{\centering\arraybackslash}p{3cm}|>{\centering\arraybackslash}p{2cm} |>{\centering\arraybackslash}p{2cm}|>{\centering\arraybackslash}p{2cm}|>{\centering\arraybackslash}p{2cm}|}
\hline
$r=0.5$, $\gamma=0.5$ & $\kappa = 0$ & $\kappa = 1$ & $\kappa = 2$ & $\kappa = 3$ \\ \cline{1-5}
{$S_{Q} (t \rightarrow \infty)$}  
 & $4.367690$ & $4.461696$ & $4.533932$ & $4.593092$  \\ \hline
$\delta_{W} (t \rightarrow \infty)$ & $0.093517$ & $0.102281$ & $0.105443$ & $0.107013$  \\ \hline
\end{tabular}
\end{center}
\caption{Asymptotic ($t \gg \gamma^{-1}$) limits of the dynamical variables ($S_{Q}, \delta_{W}$) 
for multiple photon-added states for a fixed squeezing parameter are entered following the  observations in
 Figs. \ref{fig_phase_decay} ($\mathsf{a}_{8}$) and 
($\mathsf{a}_{9}$), respectively. The parameters and the phase space variable are fixed at $\mathrm{c} = 1,\delta = 1, r= 0.5, 
\gamma = 0.5, \alpha = 2.$}
 \label{PhasePhoton}
\end{table}
\begin{figure}[H]
	\captionsetup[subfigure]{labelformat=empty}
	\subfloat[(a$_{1}$)]{\includegraphics[scale=0.62]
		{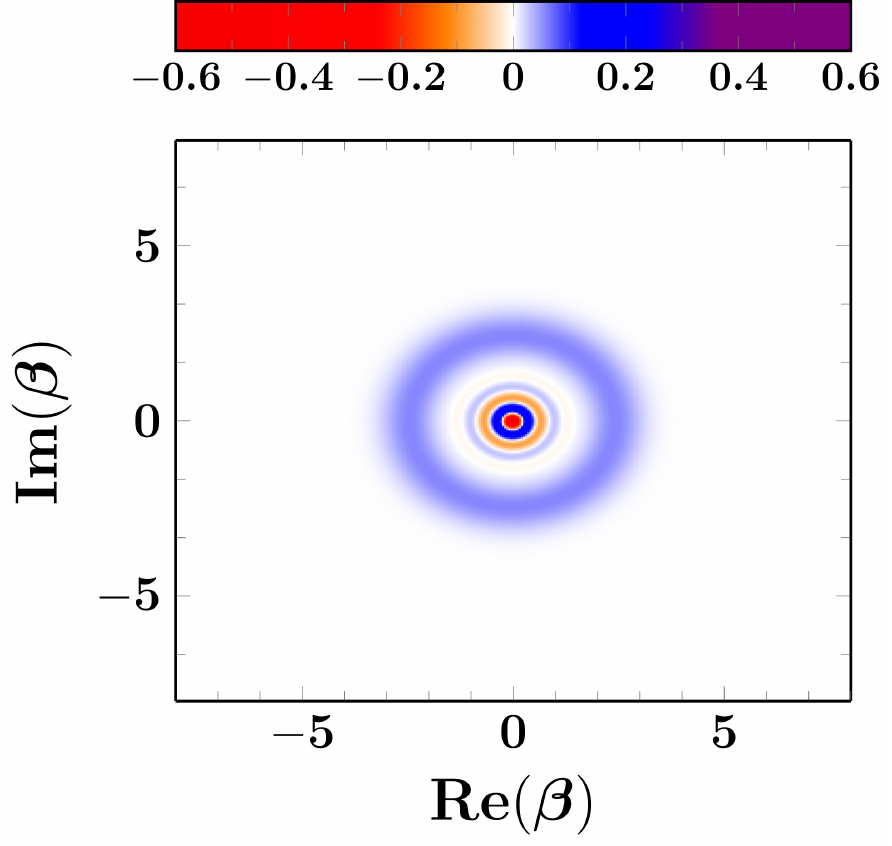}}
	\captionsetup[subfigure]{labelformat=empty}
	\subfloat[(a$_{2}$)]{\includegraphics[scale=0.61]
	{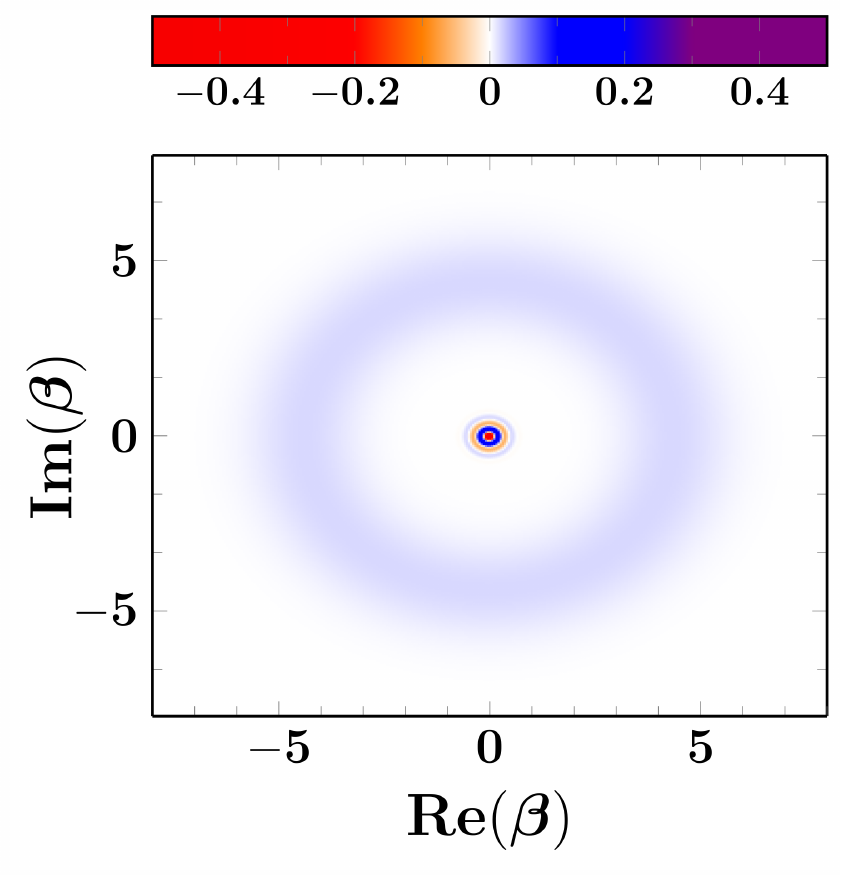}}
	\caption{The plots depict, for varying squeezing parameter $r$, the asymptotic limits of the $W$-distribution for the phase damping model  given in  (\ref{W_phase_decay}).  For the choice $\xi=0.05$ the diagram ($\mathsf{a}_{1}$) is associated with the negativity  
	 $\delta_{W}(t \rightarrow \infty)= 0.270258$, while a larger squeezing $\xi=0.7$ produces 
	($\mathsf{a}_{2}$)  a significantly lower negativity $\delta_{W}(t \rightarrow \infty)= 0.067378$, even though the phase space occupation of the latter example is much higher. The other parameters are chosen as $\mathrm{c} = 1,\kappa=1, \delta = 1, \gamma = 0.5, 
	\alpha = 2.$}
	\label{W-asymptotic}
\end{figure}
\section{Conclusion}
We have studied the evolution of a superposition of an arbitrary number of photon-added squeezed coherent cat-type states in a nonlinear 
Kerr medium. Owing to the nonlinearity of the medium the dynamical quantities such as the Wehrl entropy $S_{Q}$ and the negativity 
$\delta_{W}$ of the $W$-distribution show a periodic structure, and these quantities exhibit a series of local minima at the rational submultiples of the said period. By using the Hilbert-Schmidt distance between the quantum states we demonstrate that our evolving state  transitorily  coincides with, in general, the Yurke-Stoler type of photon-added squeezed kitten states, which maintain a uniform rotation of the phase space variables on the complex plane. For the choice of the phase space variables with macroscopically large magnitudes the kitten formations of the quantum states show extremely short-lived behavior. These transient kitten-like states allow closed form construction of the corresponding tomograms which provide the alternate description of the quantum states to the one provided by the quasiprobability distributions.  With 
the increase in number of lobes in the kitten formations, the number of interference terms increases triggering more quantumness of the          
corresponding states.  More complex quantum states embody the dominance of higher Fourier modes of the density matrix which, via rapid oscillations in the phase space, correspondingly produce more nonclassicality in the general $R$-distribution. The nonclassical depths of the states studied here are observed to attain maximum possible value. The amplitude and the phase dissipation models are studied via their related Lindblad equations. The phase damping model has the property that in the long time limit the dynamical quantities assume nontrivial asymptotic values, and the nonclassicality of the quantum states are partially retained. Even though the higher squeezing parameter is 
responsible for wider spreading of the quasiprobability functions in the phase space, it, in the long time limit, also produces less oscillatory behavior. The asymptotic limit of the negativity, therefore, decreases with increased squeezing.  
 \section*{Acknowledgement}
 One of us (VY)  acknowledges the support from DST (India) under the INSPIRE Fellowship scheme. We also acknowledge generous computational 
 help from the Department of Central Instrumentation and Service Laboratory, and the Department of Nuclear Physics, University of Madras. 
 
\end{document}